\documentclass[aps,prl,twocolumn,amsmath,amssymb,showpacs, superscriptaddress,notitlepage,longbibliography]{revtex4-1}

\usepackage[colorlinks=true,linkcolor=blue,anchorcolor=red,citecolor=blue, urlcolor=blue]{hyperref}
\usepackage{bm}
\usepackage{graphicx}
\usepackage{color}
\usepackage{cases}

\begin{document}

\title{3D quantum Hall effect of Fermi arcs in topological semimetals}

\author{C. M. Wang}

\affiliation{Institute for Quantum Science and Engineering and Department of Physics, South University of Science and Technology of China, Shenzhen 518055, China}

\affiliation{School of Physics and Electrical Engineering, Anyang Normal University, Anyang 455000, China}
\affiliation{Shenzhen Key Laboratory of Quantum Science and Engineering, Shenzhen 518055, China}

\author{Hai-Peng Sun}

\affiliation{Institute for Quantum Science and Engineering and Department of Physics, South University of Science and Technology of China, Shenzhen 518055, China}
\affiliation{Shenzhen Key Laboratory of Quantum Science and Engineering, Shenzhen 518055, China}

\author{Hai-Zhou Lu}
\email{luhz@sustc.edu.cn}
\affiliation{Institute for Quantum Science and Engineering and Department of Physics, South University of Science and Technology of China, Shenzhen 518055, China}
\affiliation{Shenzhen Key Laboratory of Quantum Science and Engineering, Shenzhen 518055, China}

\author{X. C. Xie}
\affiliation{International Center for Quantum Materials, School of Physics, Peking University, Beijing 100871, China}
\affiliation{Collaborative Innovation Center of Quantum Matter, Beijing 100871, China}

\begin{abstract}
The quantum Hall effect is usually observed in 2D systems. We show that the Fermi arcs can give rise to a distinctive 3D quantum Hall effect in topological semimetals. Because of the topological constraint, the Fermi arc at a single surface has an open Fermi surface, which cannot host the quantum Hall effect. Via a ``wormhole" tunneling assisted by the Weyl nodes, the Fermi arcs at opposite surfaces can form a complete Fermi loop and support the quantum Hall effect. The edge states of the Fermi arcs show a unique 3D distribution, giving an example of ($d$-2)-dimensional boundary states. This is distinctly different from the surface-state quantum Hall effect from a single surface of topological insulator. As the Fermi energy sweeps through the Weyl nodes, the sheet Hall conductivity evolves from the $1/B$ dependence to quantized plateaus at the Weyl nodes. This behavior can be realized by tuning gate voltages in a slab of topological semimetal, such as the TaAs family, Cd$_3$As$_2$, or Na$_3$Bi. This work will be instructive not only for searching transport signatures of the Fermi arcs but also for exploring novel electron gases in other topological phases of matter.
\end{abstract}

\date{\today}

\maketitle

\begin{figure}[htbp]
\centering
\includegraphics[width=0.9\columnwidth]{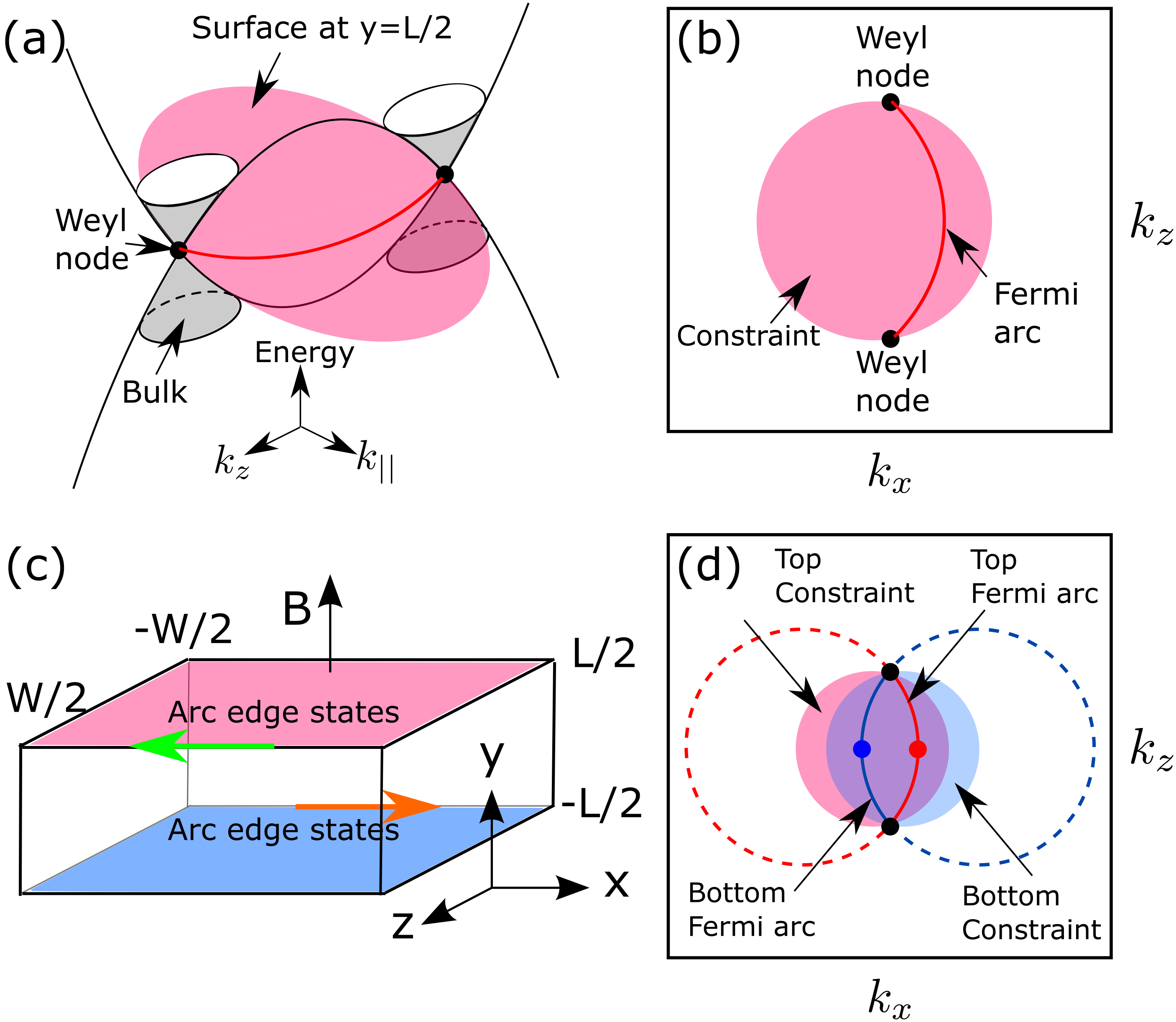}
\includegraphics[width=0.9\columnwidth]{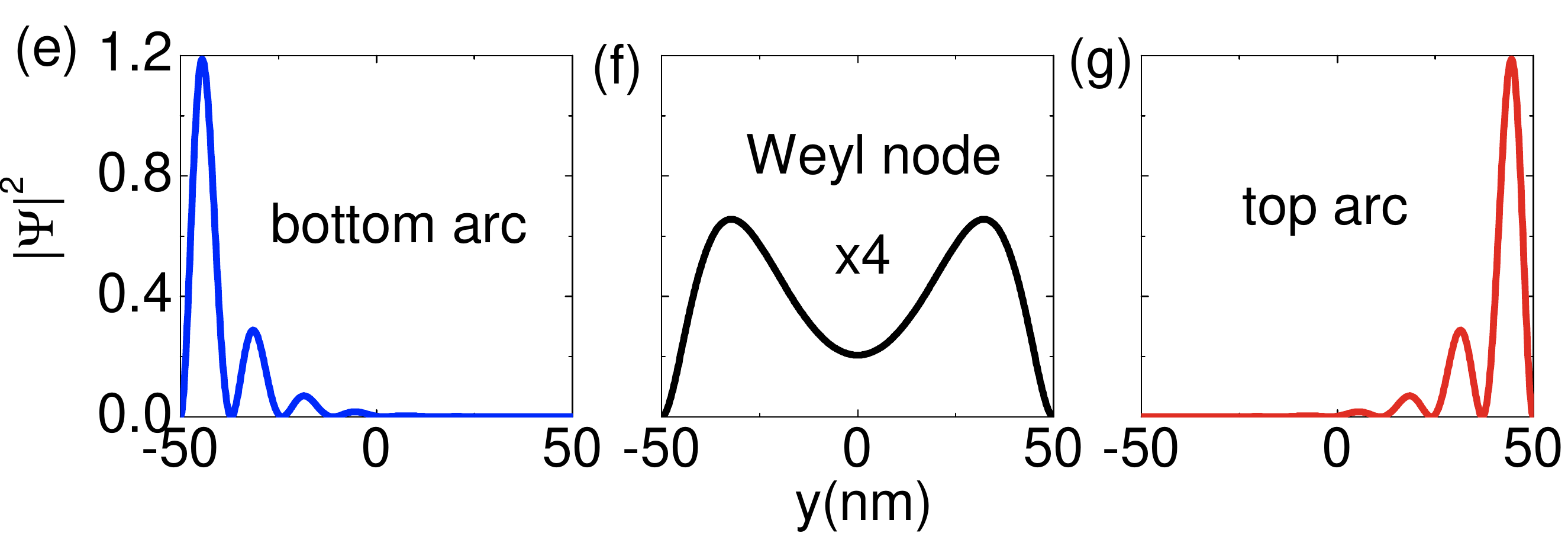}
\includegraphics[width=0.45\columnwidth]{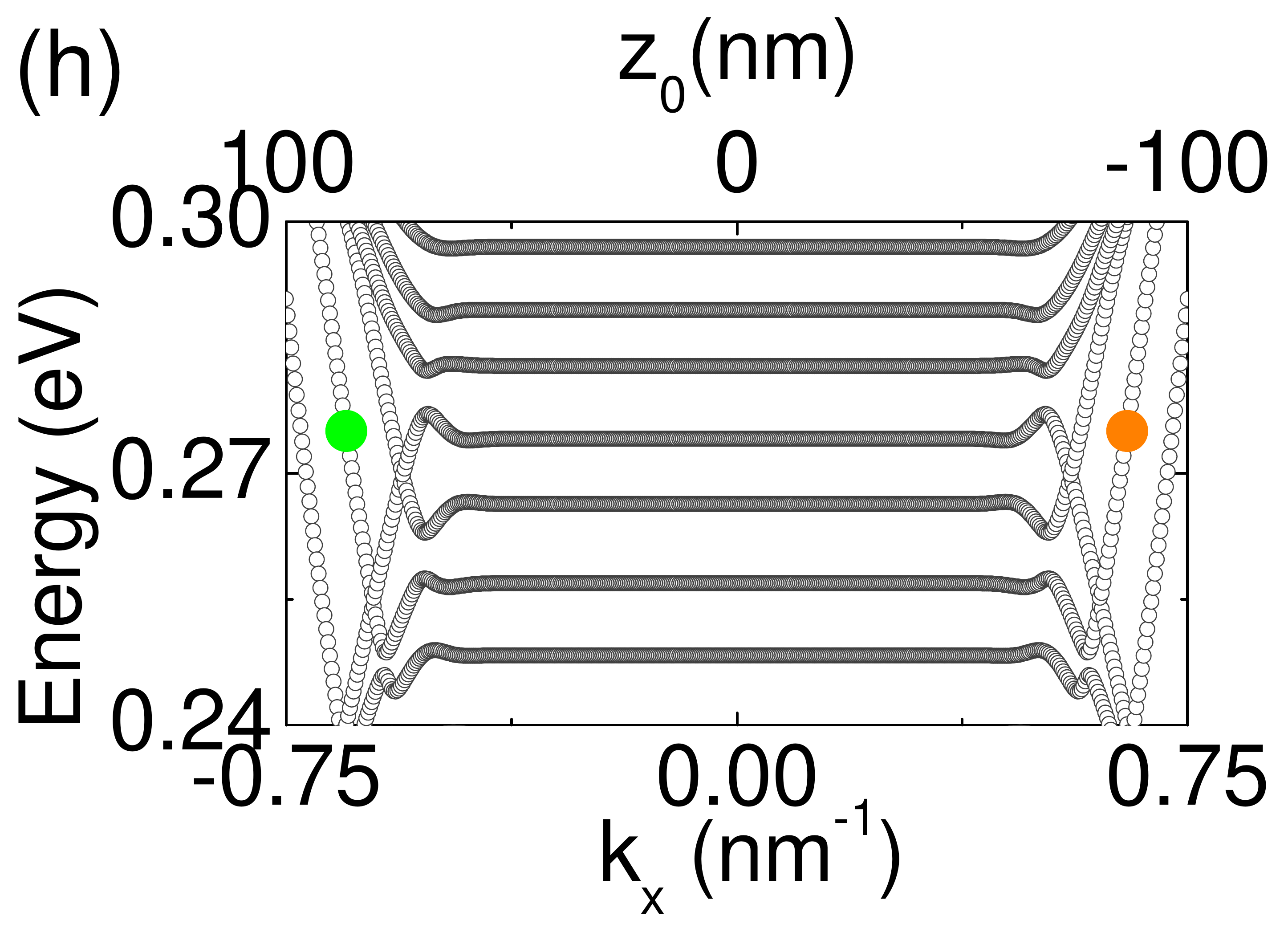}
\includegraphics[width=0.45\columnwidth]{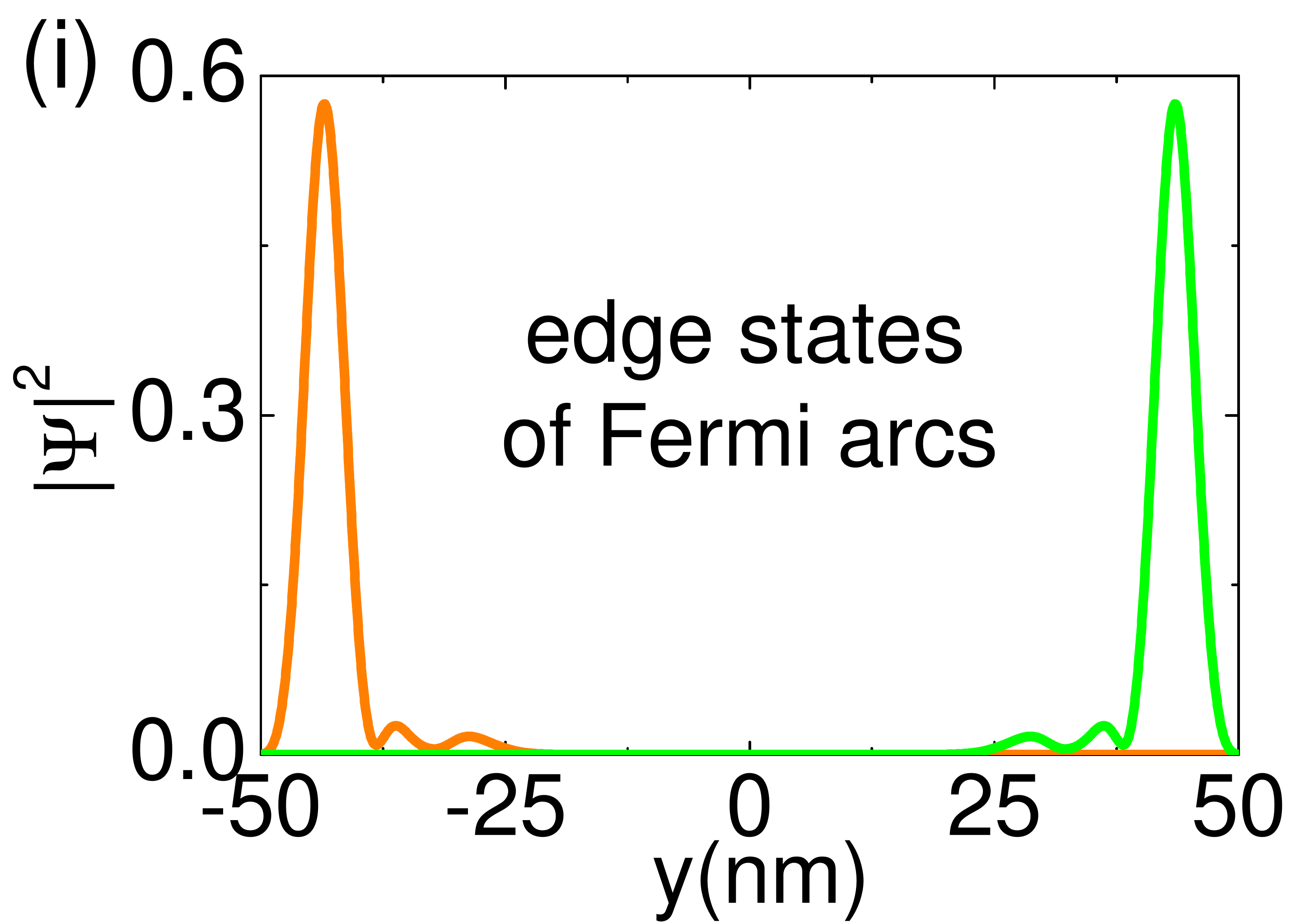}
\caption{(a) The energy dispersions for the Fermi arc (at $y=L/2$) and bulk states in a topological Weyl semimetal. $k_{||}$ stands for $(k_x,k_y)$ for the bulk and $k_x$ for the arc, respectively. (b) The Fermi arc at $y=L/2$ and $E_F=E_w$ on the $k_z-k_x$ plane. The shadow defines the ``constraint" region where the Fermi arcs can exist. (c) A slab of topological semimetal of thickness $L$ and width $W$. (d) The Fermi arcs at $E_F=E_w$ (solid) and constraints (shadow) at the $y=L/2$ (red) and $-L/2$ (blue) surfaces of the slab. [(e)-(g)] The wave function distributions at $k_z=0$ along the $y$ axis, at the blue (bottom arc, $k_x<0$), black (Weyl nodes), and red (top arc, $k_x>0$) dots in (d). (h) Landau levels of the Fermi arcs at $B=5$ T vs. the guiding center $z_0$. (i) The wave function distributions along the $y$ axis for the edge states of the Fermi arcs marked by the green and orange dots in (h). $L=100$ nm, $W=200$ nm, and other parameters can be found in Fig. \ref{Fig:Hall}.}\label{Fig:arcs}
\end{figure}

{\color{red}\emph{Introduction}} -
The discovery of the quantum Hall effect opens the door to the field of topological phases of matter \cite{Klitzing80prl}. In a strong magnetic field, the energy spectrum of a 2D electron gas evolves into Landau levels. The Landau levels deform at the sample edges and cut through the Fermi energy, forming 1D edge states protected by topology \cite{Thouless82prl}.
Electrons can flow through the edge states in a dissipationless manner, giving rise to Hall conductance in units of $e^2/h$ that defines the quantum Hall effect.
The quantum Hall effect can give transport signatures that distinguish different electron gases, such as the half-integer Hall conductance of the 2D massless Dirac fermions in graphene and topological surface states \cite{Klitzing80prl,Novoselov05nat,ZhangYB05nat,Xu14np,Yoshimi15nc}. By contrast, in a 3D electron gas, the extra dimension along the magnetic field direction prevents the quantization of the Hall conductance. Therefore, the quantum Hall effect is usually observed in 2D systems.

In this Letter, we show that the quantum Hall effect is possible in a unique 3D system, specifically, in a topological semimetal, because of the Fermi arcs. The topological semimetal is a 3D topological state of matter \cite{Wan11prb,Yang11prb,Burkov11prl,Xu11prl,Delplace12epl,Jiang12pra,Young12prl,Wang12prb,Singh12prb,Wang13prb,LiuJP14prb,Bulmash14prb}, in which energy bands touch at discrete Weyl nodes [Fig \ref{Fig:arcs} (a)].
It is equivalent to a 2D topological insulator for momenta ($k_z$ here) between the Weyl nodes, leading to the topologically protected states located at the surfaces [top and bottom in Fig. \ref{Fig:arcs} (c)] parallel to the Weyl node separation direction. The protected states form the Fermi arcs on the Fermi surface [red curves in Figs. \ref{Fig:arcs} (a) and \ref{Fig:arcs}(b)]. The Fermi arcs have been seen by ARPES  \cite{Brahlek12prl,Wu13natphys,Wang12prb,Liu14sci,Wang13prb,Xu15sci,Wang13prb,Liu14natmat,Neupane14nc,Yi14srep,Borisenko14prl,Weng15prx,Huang15nc,Lv15prx,Xu15sci-TaAs}
and can induce novel quantum oscillations \cite{Potter14nc,Moll16nat}.
Topological phases of matter usually come with distinctive transports, making the transport signature of the Fermi arcs an intriguing topic \cite{Hosur12prb,Baum15prx,Gorbar16prb,Ominato16prb,McCormick17arXiv}.

There are several issues for the Fermi arcs to exhibit the quantum Hall effect. First, the topological origin requires that the states of Fermi arcs occupy only a region between the Weyl nodes \cite{ZhangSB16njp} [Fig. \ref{Fig:arcs}(b)]. At one surface, the Fermi arcs cannot form a closed Fermi loop needed by Landau levels and the quantum Hall effect. We find that the Fermi arcs from opposite surfaces in a topological semimetal slab [Fig. \ref{Fig:arcs}(c)] can complete the needed closed Fermi loop [Fig. \ref{Fig:arcs}(d)]. Electrons can tunnel between the Fermi arcs at opposite surfaces via the Weyl nodes [Figs. \ref{Fig:arcs}(e)-\ref{Fig:arcs}(g)]. Second, the quantum Hall effect solely from the Fermi arcs requires the bulk carriers to be depleted by tuning the Fermi energy to the Weyl nodes \cite{Ruan16nc}. Third, we find that the band anisotropy in the bulk Weyl fermions is necessary for the Fermi arcs to form a 2D electron gas. These properties in the quantum Hall effect can provide transport signatures for the Fermi arcs. Compared to the novel quantum oscillations \cite{Potter14nc,Moll16nat}, the quantum Hall effect of the Fermi arcs contributes a quantum complement to the Fermi arc dominant electronic transports.
The Weyl semimetals TaAs family \cite{Weng15prx,Huang15nc,Lv15prx,Xu15sci-TaAs,HuangXC15prx,Yang15np,Shekhar15np,ZhangCL16nc} and the Dirac semimetals Cd$_3$As$_2$ and Na$_3$Bi have extremely high mobilities \cite{He14prl,Liang15nmat,Zhao15prx,Narayanan15prl,Xiong15sci} required by the quantum Hall effect. Low carrier densities \cite{LiCZ15nc,LiH16nc,ZhangC17nc} and gating \cite{LiCZ15nc} have also been achieved. We expect the quantum Hall effect of the Fermi arcs in slabs of the TaAs family, [110] or [1$\bar{1}$0] Cd$_3$As$_2$ \cite{Uchida17MarchMeeting}, and [100] or [010] Na$_3$Bi.

{\color{red}\emph{Minimal model}} - We will use a minimal model to illustrate the physics for  the Fermi-arc quantum Hall effect. To preserve their topological properties, we need to derive the 2D effective model of the Fermi arcs from a 3D model of Weyl semimetal \cite{Shen12book,Okugawa14prb,Lu15Weyl-shortrange}
\begin{equation}\label{Ham}
 H=D_1k_y^2+D_2(k_x^2+k_z^2)+A(k_x\sigma_x+k_y\sigma_y)+M(k_w^2-k^2)\sigma_z,
\end{equation}
where $ (\sigma_x,\sigma_y,\sigma_z)$ are the Pauli matrices, the wave vector $\mathbf{k}=(k_x,k_y,k_z)$, and $D_1$, $D_2$, $A$, $M$, and $k_w$ are model parameters. We assume that $ |M|>|D_1|$. The energy dispersion of the model is $E_{\pm }^{\mathbf{k}}=D_1k_y^2+D_2(k_x^2+k_z^2)\pm [M^2(k_w^2-\mathbf{k}^2)^2 +A^2(k_x^2+k_y^2)]^{1/2}$, with $\pm$ for the conduction and valence bands, respectively. The model hosts two Weyl nodes at $(0,0,\pm k_w)$ having energy $E_w=D_2k_w^2$ [Fig. \ref{Fig:arcs} (a)], and carries all of the topological semimetal properties \cite{Lu17fop}. In contrast to the $k\cdot \sigma$ model, the Fermi arc states can be solved analytically from the model \cite{ZhangSB16njp}.

{\color{red}\emph{Open Fermi arc at one surface}} - First, we show that the Fermi arc at a single surface of a Weyl semimetal cannot host the quantum Hall effect. We focus on the $y=L/2$ surface. By replacing $k_y$ with $-i\partial_y $ in $H$ and using open-boundary conditions, we can solve the wave function at $k_x=k_z=0$, and then project $H$ on the wave function to construct the effective model (see the procedure at \cite{Lu10prb,Shan11njp,ZhangSB16njp} and Sec. S1 of \cite{Supp}) for the Fermi arc
\begin{eqnarray}\label{Fermi_arc}
   H_{\rm arc}=D_1k_w^2+ v k_x +(D_2-D_1)(k_x^2+k_z^2),
\end{eqnarray}
where $ v\equiv A\sqrt{M^2-D_1^2}/M$.
If there is no anisotropic $D$ terms, the Fermi arc only disperses linearly with $k_x$; consequently, the Landau levels cannot be defined. Therefore, the anisotropic $D$ terms are necessary. Moreover, the electron gas of the Fermi arc is distinct from usual 2D electron gases because it is confined within a specific momentum space due to their topological nature \cite{ZhangSB16njp}. For this model, the Fermi arc at the $y=L/2$ surface is confined in a region defined by the constraint
\begin{eqnarray}\label{cons}
k_x^2+k_z^2+2a k_x <k_w^2,
\end{eqnarray}
where $a\equiv AD_1/2M\sqrt{M^2-D_1^2}$. This means that, the wave vectors of the Fermi arcs at the $y=L/2$ surface are only allowed within a circle of radius $\sqrt{k_w^2+a^2}$ centered at $(k_x$=$-a$, $k_z$=$0)$. The Fermi circle of $H_{\rm arc}$ at a given Fermi energy can only partially overlap with the constraint in Eq. (\ref{cons}), forming an ``open'' Fermi surface, as shown by the red solid curve in Figs. \ref{Fig:arcs}(a) and \ref{Fig:arcs}(b).
Because of the open Fermi surface, electrons cannot undergo complete cyclotron motion in a perpendicular magnetic field. Thus, the 2D electron gas of Fermi arc at a single surface cannot form well-defined Landau levels required by the quantum Hall effect.

\begin{figure}
\centering
\includegraphics[width=0.49\columnwidth]{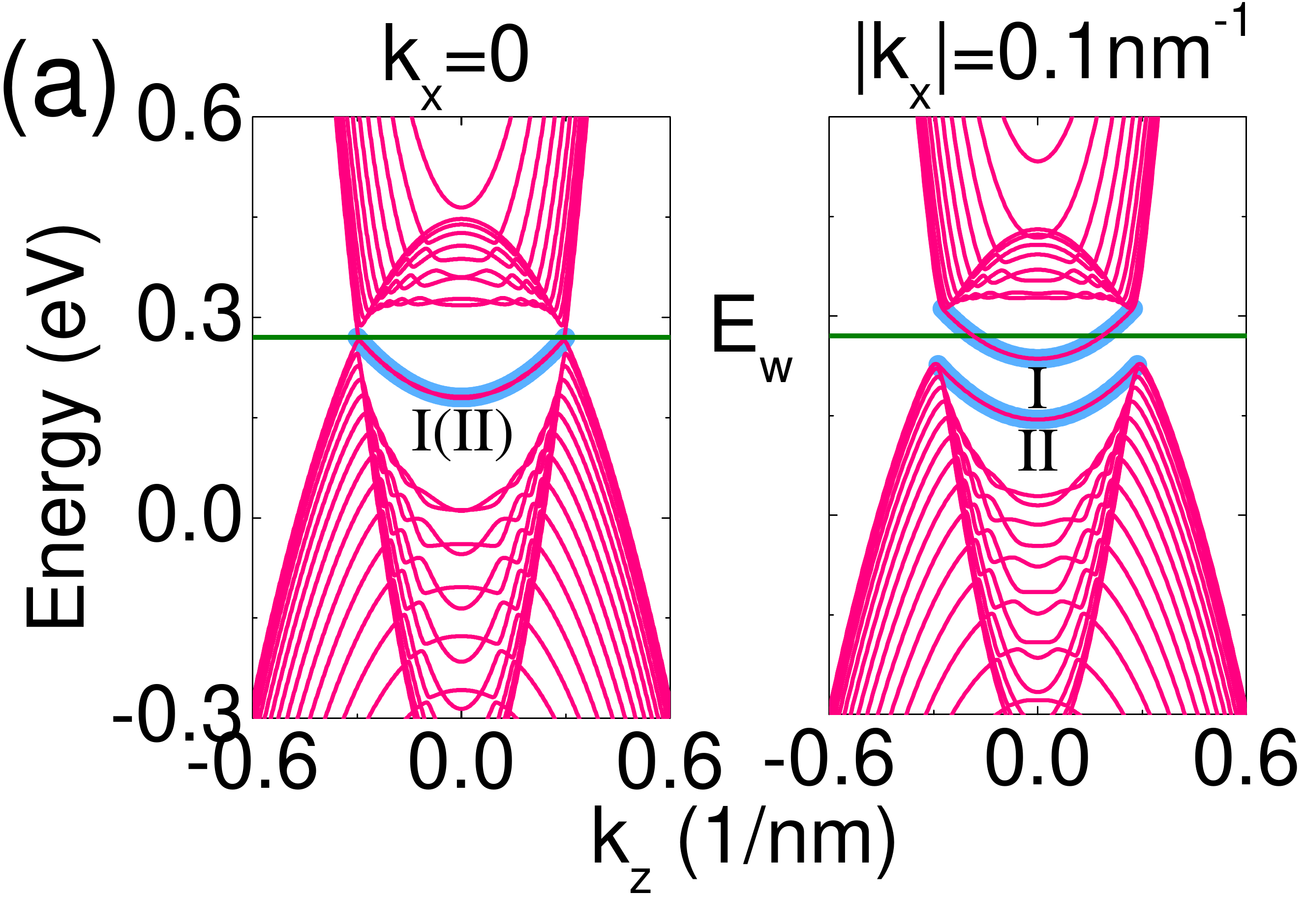}
\includegraphics[width=0.49\columnwidth]{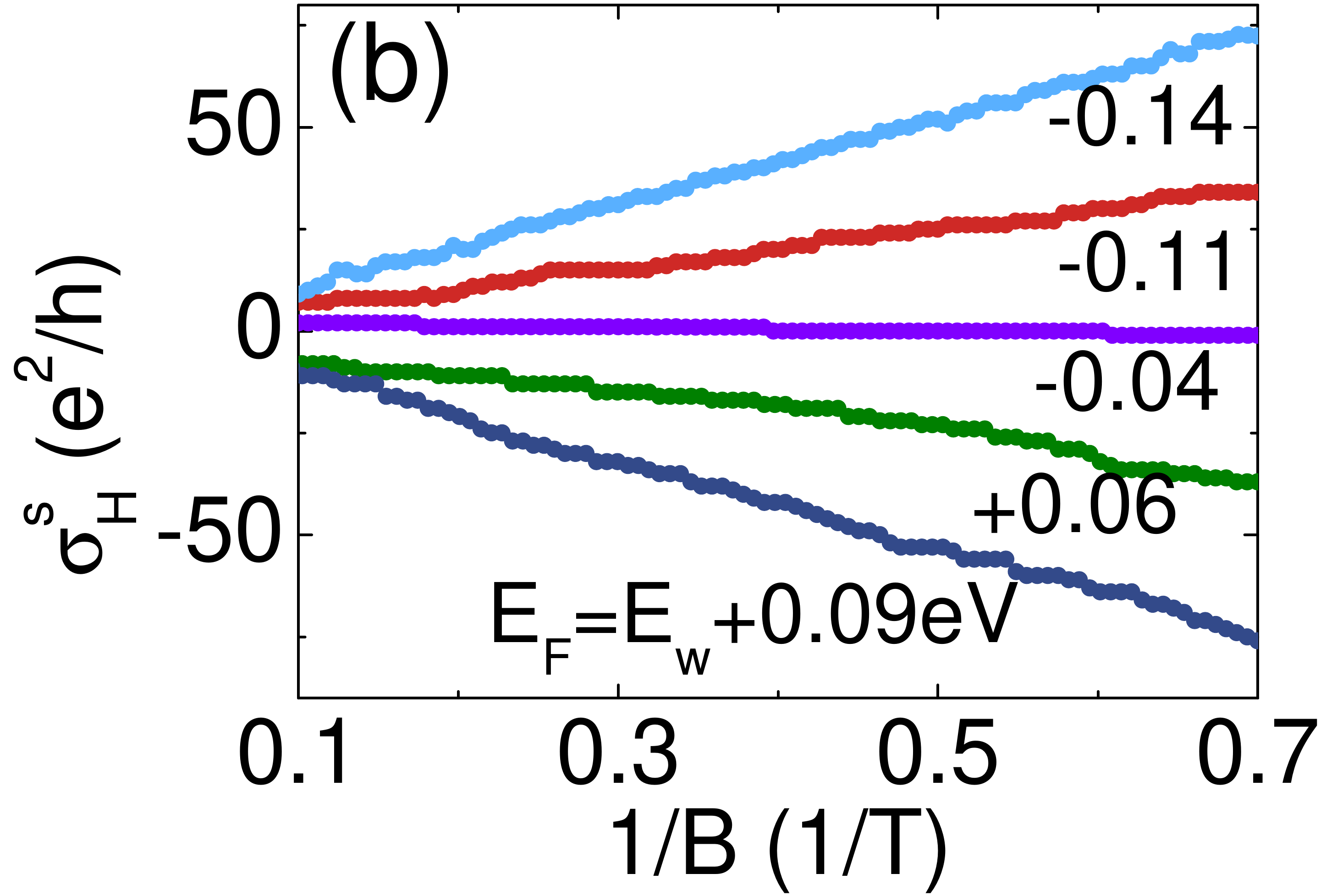}
\includegraphics[width=0.95\columnwidth]{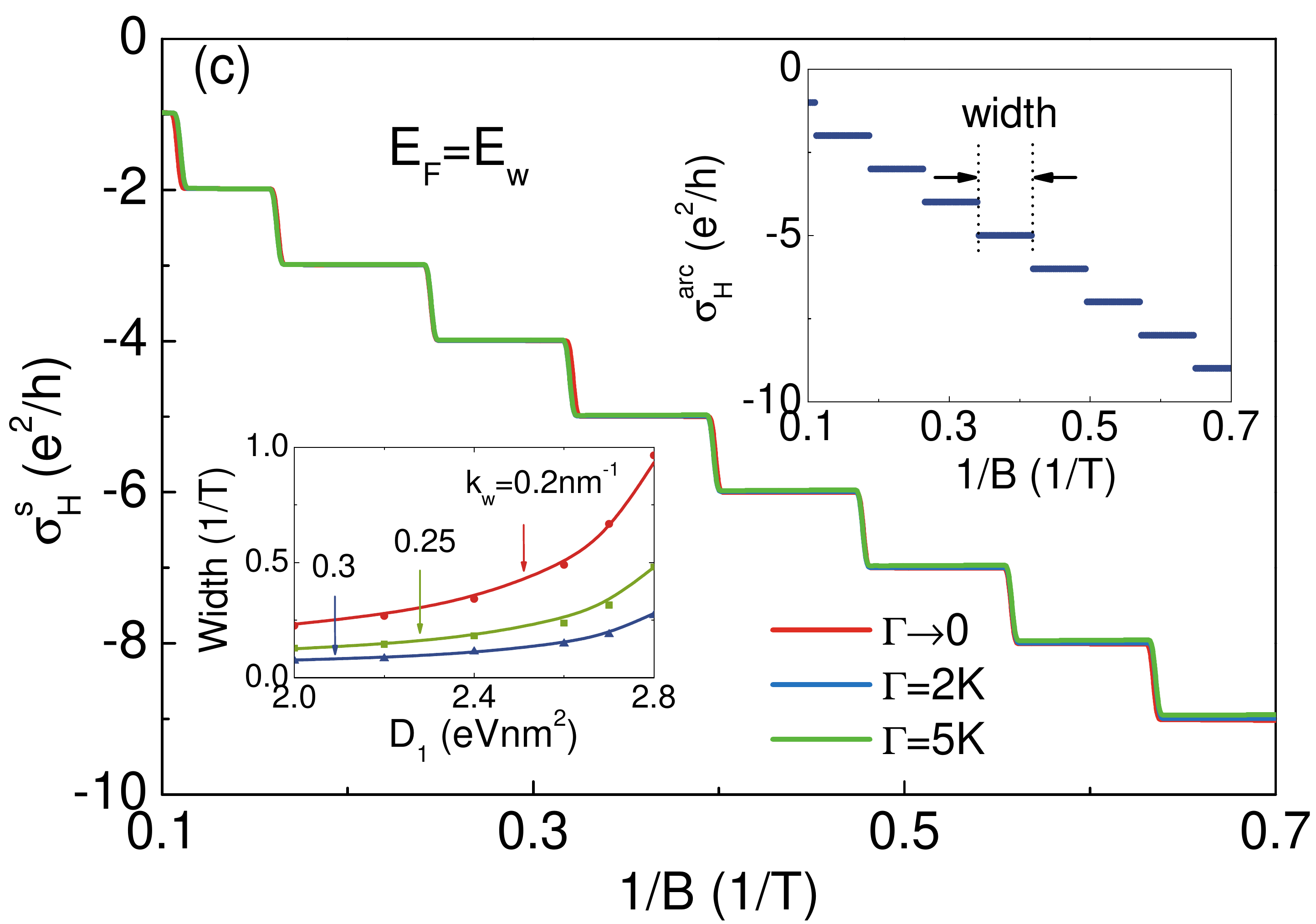}
\caption{(a) In a topological semimetal slab, the numerically calculated energy spectrum (pink) for the bulk states and Fermi arcs at $k_x=0$ (left) and $k_x=\pm 0.1$ nm$^{-1}$ (right). The blue curves are the Fermi arc bands plotted using $H_{\textrm{arc}}$ and $H_{\textrm{arc}}'$. (b) The sheet Hall conductivity when the Fermi energy $E_F$ crosses the bulk states for $\Gamma\rightarrow 0$. $\Gamma$ is the disorder-induced level broadening. Recent experiments show that gating can tune carriers from $n$- to $p$-type in 100-nm-thick devices of topological semimetal \cite{LiCZ15nc}. (c) The sheet Hall conductivity $\sigma_{\rm H}^s$ at $E_F=E_w$, where the Fermi energy crosses only arc I. The right inset shows the analytic Hall conductance $\sigma_{\rm H}$ in Eq. \eqref{analytical}. In the presence of a residual detuning from the Weyl nodes, the bulk states also contribute to $\sigma_{\rm H}^s$. Unlike that from the Fermi arcs, the contribution from the bulk states may change with the slab thickness. The left inset shows the width of the Hall plateaus in the clean limit as a function of $D_1$ for different $k_w$. The dots and lines are the numerical and analytic results, respectively. The parameters are $M$=$5$eVnm$^2$, $A$=0.5eVnm, and $D_2$=$3$eVnm$^2$, $D_1$=$2$eVnm$^2$, $k_w$=$0.3$nm$^{-1}$, and $L$=$100$ nm.}\label{Fig:Hall}
\end{figure}

{\color{red}\emph{Fermi arc loop via ``wormhole" tunneling}} -
In contrast, the Fermi arcs at two opposite surfaces of a slab of Weyl semimetal,  with the assistance of the Weyl nodes, can form a closed Fermi loop to support the quantum Hall effect. For a Weyl semimetal slab of thickness $L$, we consider two opposite surfaces at $y=\pm L/2$ [Fig.~\ref{Fig:arcs} (c)]. Similar to Eqs.~\eqref{Fermi_arc} and \eqref{cons}, the model and constraint at the $y=-L/2$ surface are found as
$ H'_{\rm arc}=D_1k_w^2- v k_x +(D_2-D_1)(k_x^2+k_z^2)$ and
$k_x^2+k_z^2-2ak_x<k_w^2$,
respectively. Figure \ref{Fig:arcs} (d) shows the Fermi arcs at $E_F=E_w$ and constraints at the two surfaces. The Fermi arcs at opposite surfaces shift along opposite directions on the $k_x$ axis. The two open Fermi arcs [red and blue curves in Fig. \ref{Fig:arcs} (d)] can form a Fermi loop well inside the overlapping constraint regions; thus, all states on this loop are allowed. We numerically calculate the energy spectrum for this slab by using the basis $ \varphi_n(y)=\sqrt{2/L}\sin [ n\pi (y/L+1/2 )]$ (Sec. S2 of \cite{Supp}). Figure \ref{Fig:Hall} (a) verifies the above picture for the Fermi loop formation. The energy band for the Fermi loop is marked as ``I" (arc I). There is another band (marked as ``II"), which appears below arc I at $k_x=\pm 0.1 $ nm $^{-1}$ but buried in the bulk valence bands.
Moreover, the wave function on the Fermi loop can evolve from located at one surface [Figs. \ref{Fig:arcs}(e) and \ref{Fig:arcs}(g)] to spread out in the $y$ direction [Fig. \ref{Fig:arcs} (f)] when moving from the Fermi arcs to the Weyl nodes. Therefore, the Weyl nodes act like ``wormholes" that connect the top and bottom surfaces, and an electron can complete the cyclotron motion. Because the Weyl nodes are singularities in both energy and momentum, the wormhole tunneling can be infinite in both time and space, according to the uncertainty principle.
In realistic materials, the tunneling distance is limited by the mean free path, which can be comparable to or longer than 100 nm in high-mobility topological semimetals \cite{HuangXC15prx,Yang15np,Shekhar15np,ZhangCL16nc,He14prl,Liang15nmat,Zhao15prx,Narayanan15prl,Xiong15sci}, even up to 1 $\mu$m \cite{Moll16nat}, so the thickness in the calculation is chosen to be 100 nm. The loop formed by the Fermi arcs at opposite surfaces via the Weyl nodes can support a 3D quantum Hall effect. The wormhole effect has been addressed in different situations in topological insulators \cite{Rosenberg10prb}.

{\color{red}\emph{The Hall response}} - Now we demonstrate that arc I of the Weyl semimetal slab can host the quantum Hall effect. The Hall conductivity can be calculated from the Kubo formula (Sec. S3 of \cite{Supp})
\begin{equation}\label{Kubo}
  \sigma_{\rm H}=\frac{e^2\hbar}{iV_{\rm eff}}\sum_{\delta,\delta'\ne\delta }\frac{\langle \Psi_\delta|v_x|\Psi_{\delta'}\rangle \langle \Psi_{\delta'}|v_z|\Psi_{\delta}\rangle [f(E_{\delta'}) -f(E_{\delta})]}{(E_{\delta}-E_{\delta'})(E_{\delta}-E_{\delta'}+i\Gamma)}.
\end{equation}
Here $|\Psi_{\delta}\rangle$ is the eigenstate of energy $E_{\delta}$ for $H$ in a $y$-direction magnetic field and with open boundaries at $y=\pm L/2$, $v_x$ and $v_z$ are the velocity operators, $f(x)$ is the Fermi distribution, $V_{\rm eff}$ is the volume of the slab or the area of the surfaces that host the Fermi arcs. $\sigma_{\rm H}$ has a dimension of $e^2/h$ in 2D and of $e^2/h$ over length in 3D. The sheet Hall conductivity for the slab can be defined as $\sigma_{\rm H}^s=\sigma_{\rm H}L$.
We use the basis $|\phi_\nu(z)\rangle \otimes |\varphi_n(y)\rangle$ to find the eigenenergies for a slab in the $y$-direction magnetic field, where $\phi_\nu$ are the harmonic oscillator eigenfunctions. Figure \ref{Fig:Hall} (b) shows the sheet Hall conductivity for the topological semimetal slab at Fermi energies far away from the Weyl nodes. $\sigma_{\rm H}^s$ follows the usual $1/B$ dependence.
As the Fermi energy is shifted towards the Weyl nodes, the
slope becomes smaller, indicating decreasing carrier density. Also, quantized plateaus of $\sigma_{\rm H}^s$
start to emerge as the Fermi energy approaches the Weyl nodes.
When the Fermi energy crosses only arc I [Fig. \ref{Fig:Hall}(c)], $\sigma_{\rm H}^s$ shows well-formed
quantized plateaus in units of $e^2/h$, indicating the quantum Hall effect of the Fermi arcs. Here disorder is included in the Kubo formula via the level broadening $\Gamma$. This treatment is capable of giving the quantization in graphene \cite{Gusynin05prl}, which is massless in 2D. Because of the relation with the Chern number \cite{Thouless82prl}, the quantum Hall effect can be theoretically studied in the absence of disorder, as those in topological insulators \cite{Zyuzin11prb,ZhangSB14prb,ZhangSB15srep,Pertsova16prb}.
To verify the numerical result in Fig. \ref{Fig:Hall}(c), we also calculate analytically the quantum Hall conductance from arc I (Sec. S4 of \cite{Supp}), by modeling arc I as an anisotropic parabolic band $ H_{\rm arc_I}\approx D_1k_w^2+\hbar^2k_x^2/2m_x + \hbar^2k_z^2/2m_z $, with $m_x$ and $m_z$ being the effective masses. We can find the quantum Hall conductance of arc I in the clean limit $\Gamma\rightarrow0$ \cite{Jain07book,Laughlin81prb}
\begin{eqnarray}
  \sigma_{\rm H}^{\mathrm{arc}}=\frac{e^2}{h}\text{sgn}(R)\text{sgn}(eB)\left\lfloor\frac{S_{\rm I}/ (2\pi)^2}{eB/h}
  +\frac{1}{2}\right\rfloor,\label{analytical}
\end{eqnarray}
where $\lfloor ...\rfloor$ stands for rounding down, $R=D_2-D_1$ and the area of arc I in momentum space is $S_{\rm I}= 2k_w^2 (1+ v^2/4R^2k_w^2) \arctan ( 2|R|k_w/|v|) -|v|k_w/|R|$. Figure \ref{Fig:Hall} (c) shows a good agreement between the analytic and numerical results on the Hall conductance and width of the plateaus.

\begin{figure}[htbp]
\centering
\includegraphics[width=0.85\columnwidth]{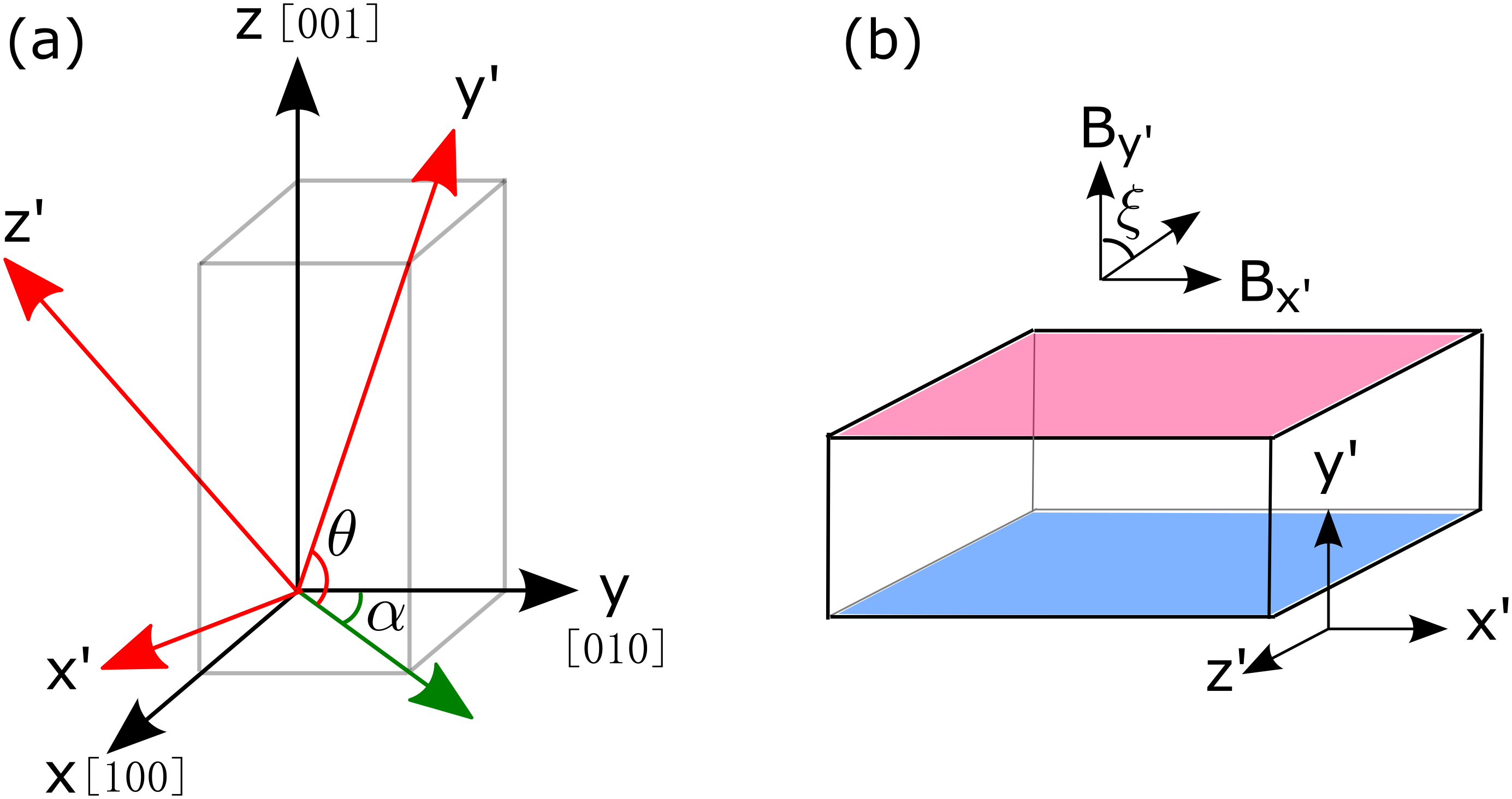}\\
\includegraphics[width=0.45\columnwidth]{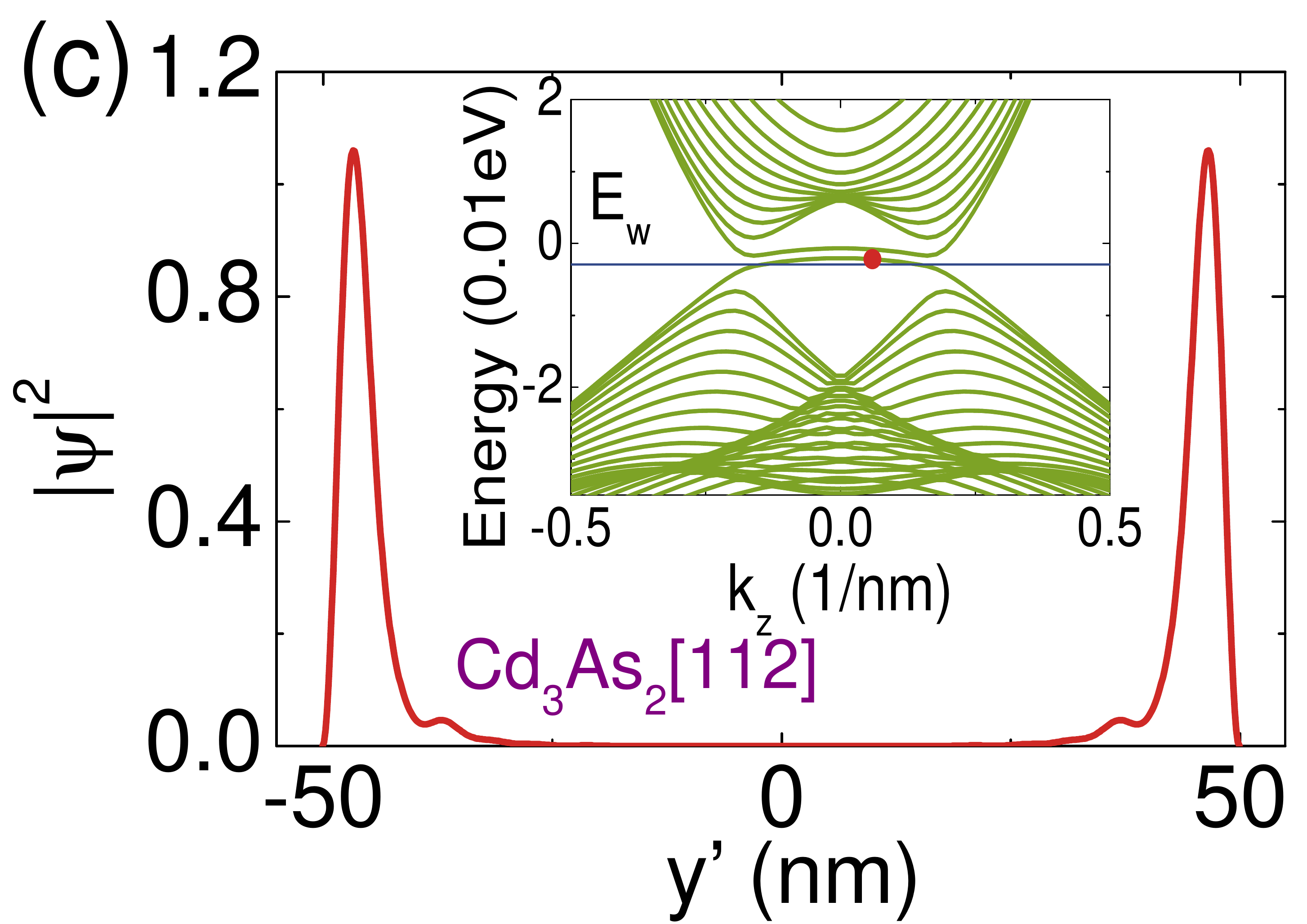}
\includegraphics[width=0.45\columnwidth]{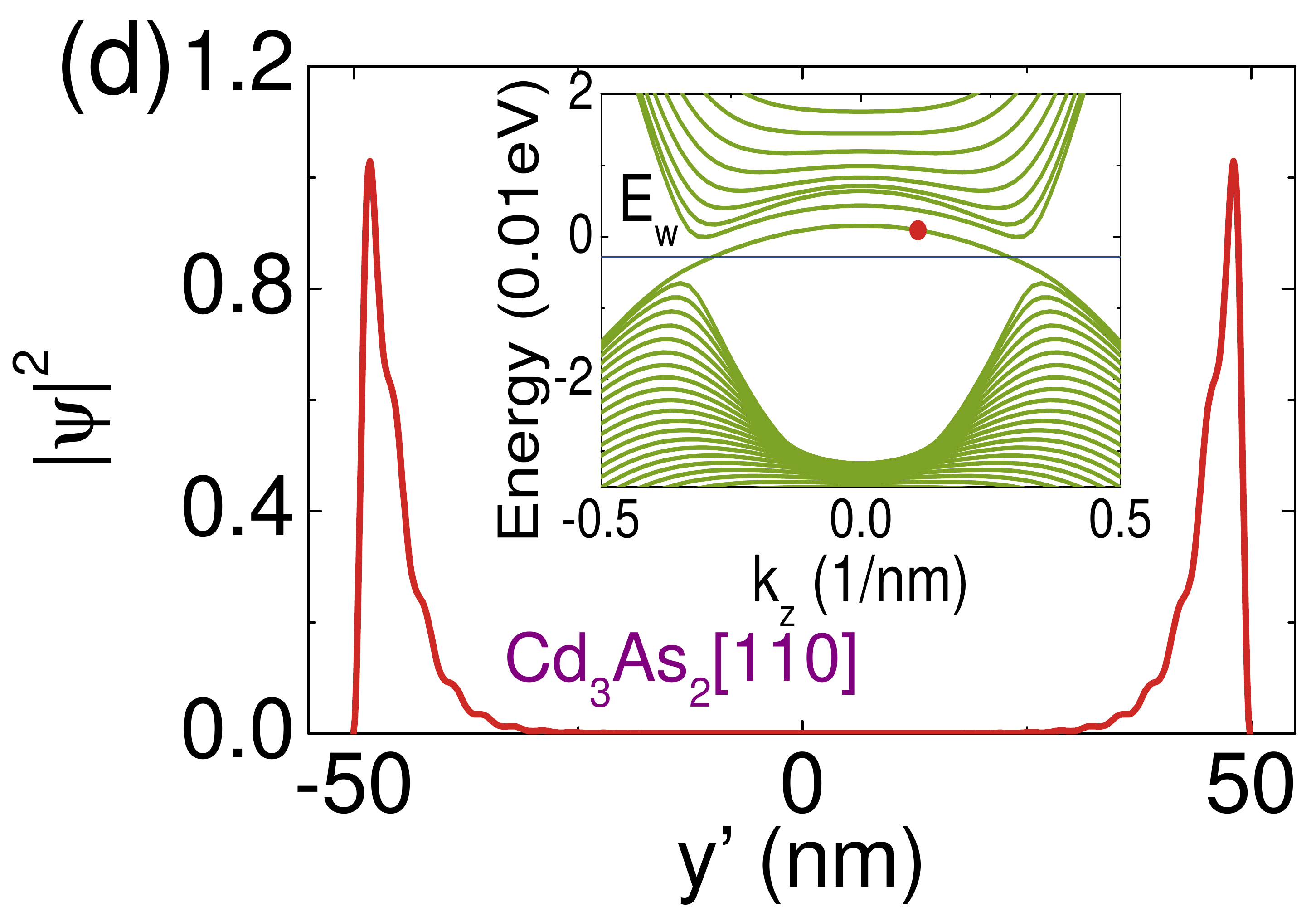}
\includegraphics[width=0.8\columnwidth]{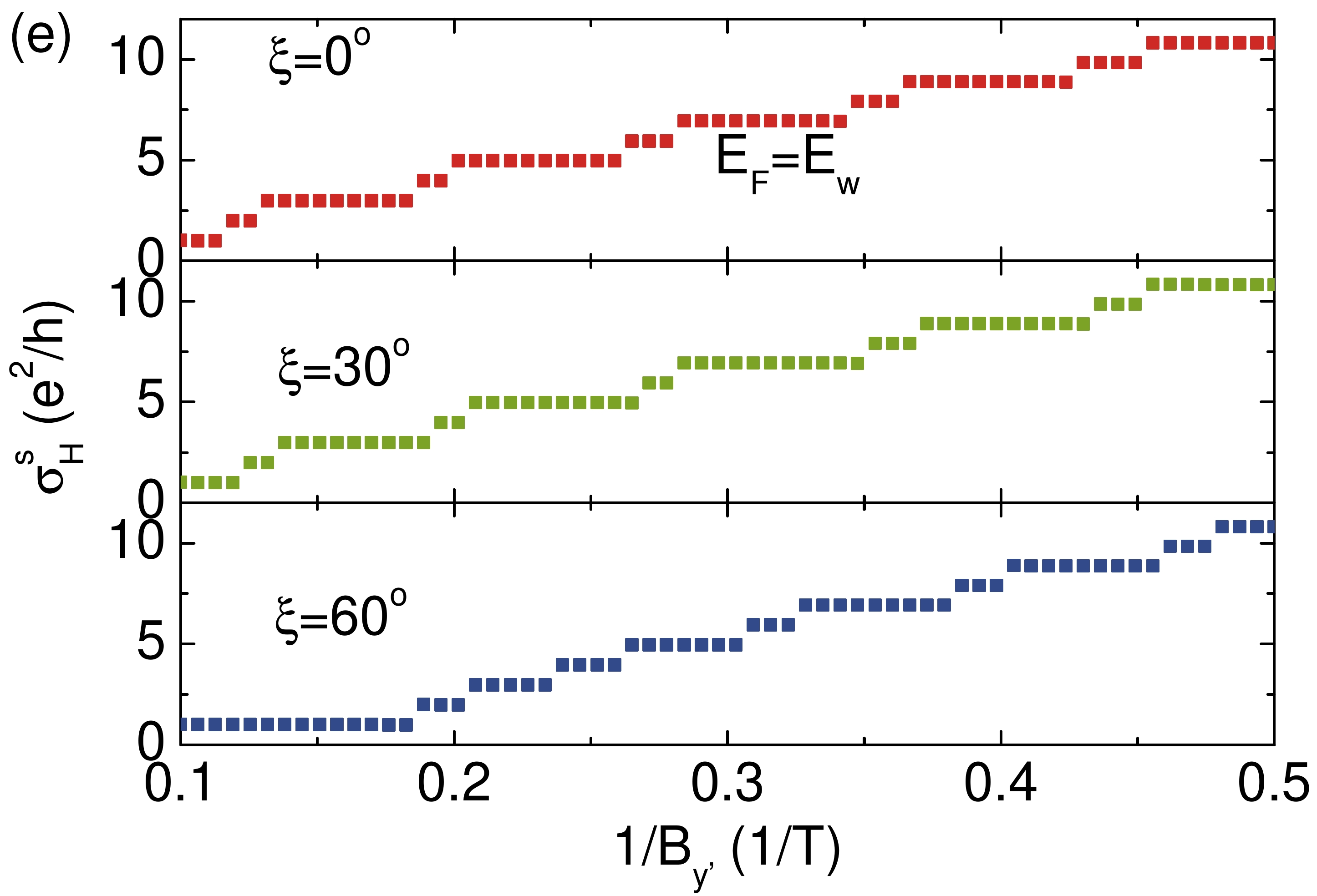}
\includegraphics[width=0.45\columnwidth]{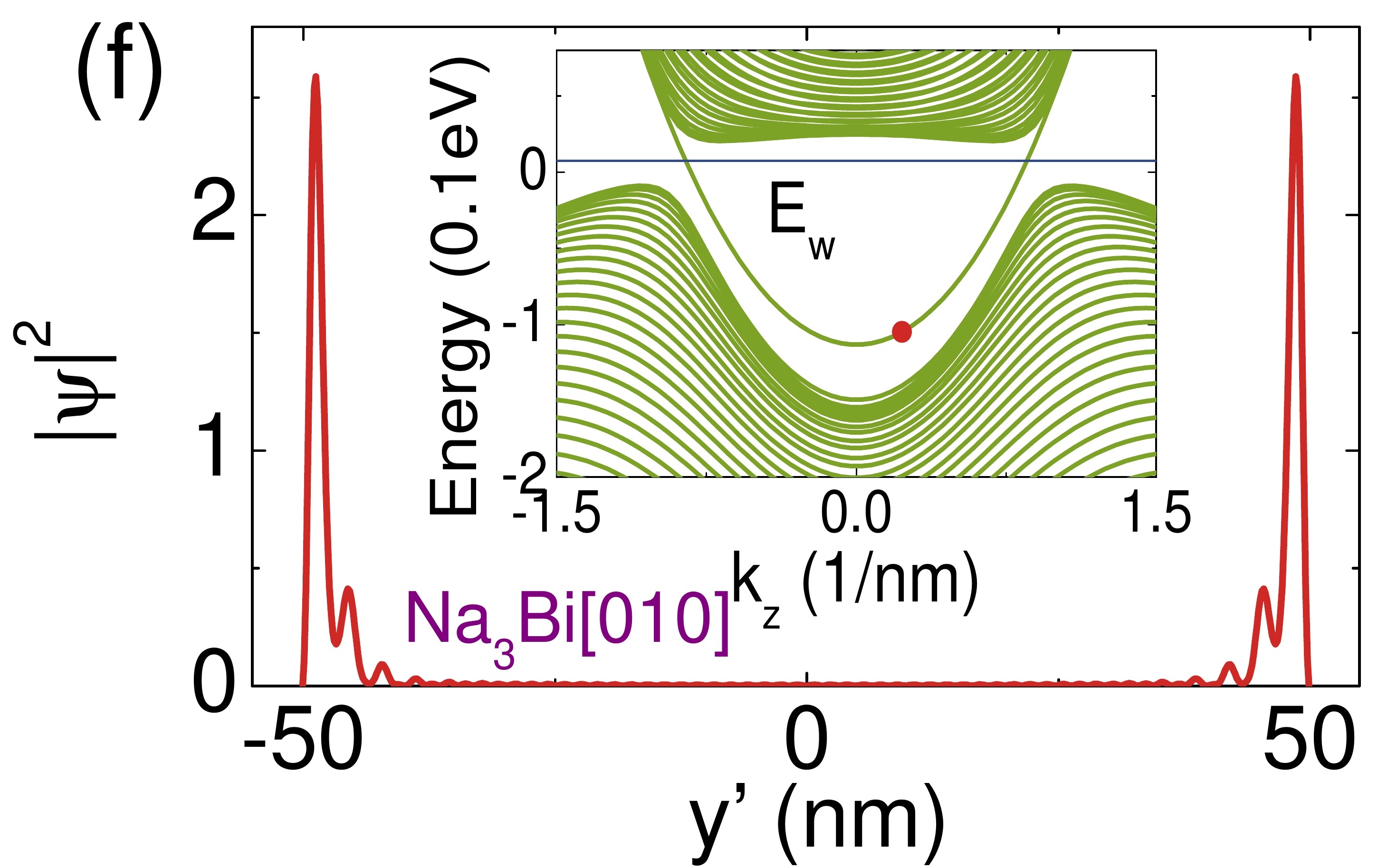}
\includegraphics[width=0.45\columnwidth]{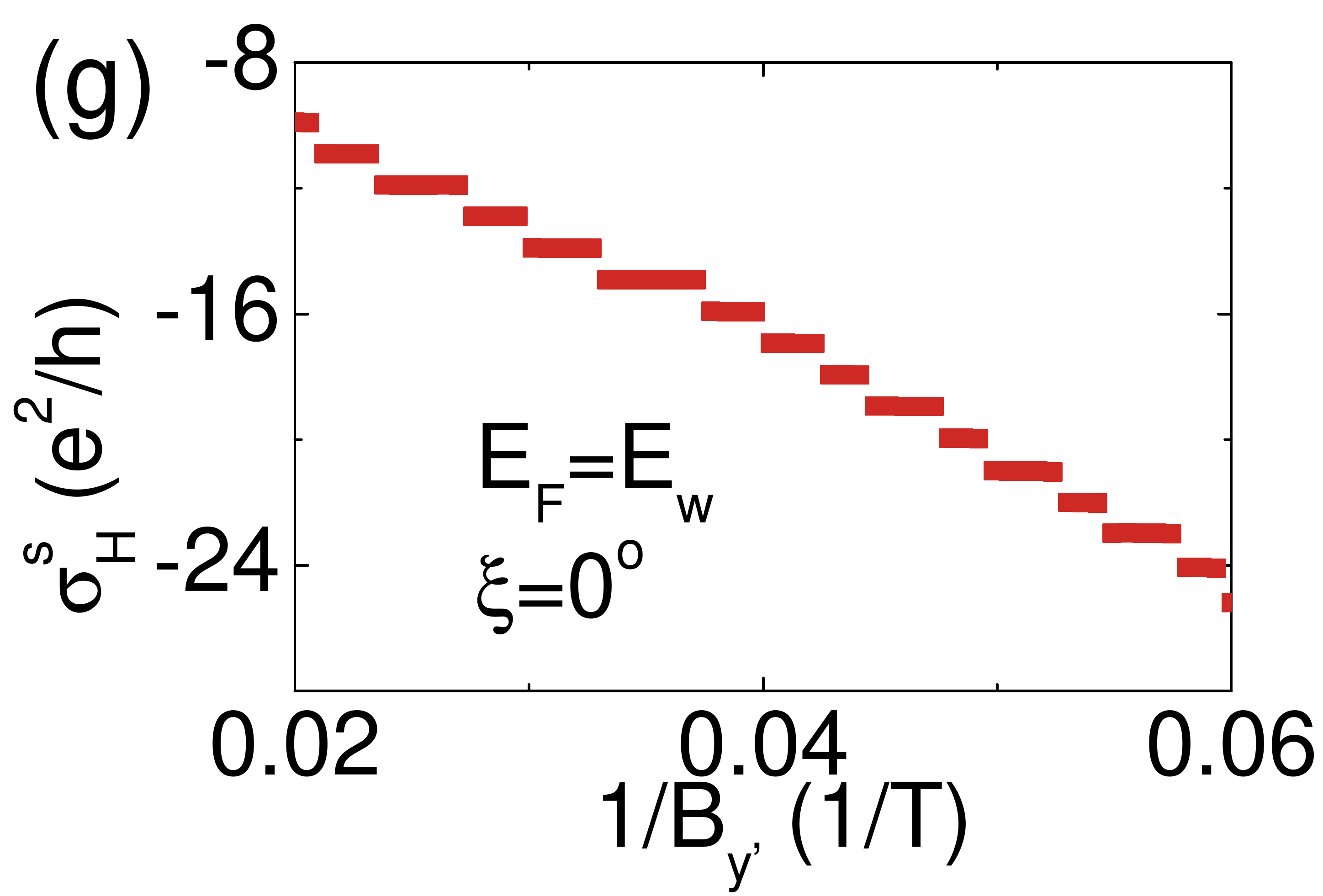}
\caption{(a) The crystallographic directions and coordinates system in the model of Dirac semimetal Eq. (\ref{Eq:H-Wang}). Slabs grown along the [112] and [110] directions corresponds to $(\alpha,\theta)=(-\pi/4,\tan^{-1}\sqrt{2})$ and $(-\pi/4, 0)$, respectively.
(b) $y'$ is the slab growth direction. The Hall conductance is defined in the $x'-z'$ plane. The magnetic field can be rotated from the $y'$ to the $x'$ direction by the angle $\xi$. [(c),(d), (f)] For a slab of 100 nm, the energy spectrum at $k_x=0.01$nm$^{-1}$ and the wave function distribution along the $y'$ direction for the states marked by the red points. (e) The sheet Hall conductivity at the Dirac node $E_w=C_0-C_1M_0/M_1$ for the Cd$_3$As$_2$ [110] slab at different $\xi$. (g) The same as (e) but for the [010] Na$_3$Bi slab. The parameters for Cd$_3$As$_2$ are $C_0$=$-0.0145$eV, $C_1$=$10.59$eV$\AA^2$, $C_2$=$11.5$eV$\AA^2$, $M_0$=$0.0205$eV, $M_1$=$-18.77$eV$\AA^2$, $M_2$=$-13.5$eV$\AA^2$, $A$=$0.889$eV$\AA$ \cite{Cano17prbrc}, $g_s$=$18.6$, and $g_p$=$2$ \cite{Jeon14nmat}. The parameters for Na$_3$Bi are $C_0$=$-0.06382$eV,  $C_1$=$8.7536$eV$\AA^2$, $C_2$=$-8.4008$eV$\AA^2$, $M_0$=$0.08686$eV, $M_1$=$-10.6424$eV$\AA^2$, $M_2$=$-10.361$eV$\AA^2$, $A$=$2.4598$eV$\AA$ \cite{Cano17prbrc} $g_s$=$20$, and $g_p$=$20$ \cite{Xiong15sci}. $\Gamma$=$1$K in panels (e) and (g).}\label{Fig:Dirac}
\end{figure}

{\color{red} \emph{Where are the edge states of the Fermi arcs?}} - Figures \ref{Fig:arcs}(h) and \ref{Fig:arcs}(i) show that the edge states of the Fermi arcs have a unique 3D  spatial distribution. Figure \ref{Fig:arcs} (h) shows the energies of the Landau levels in the $y$-direction field. The energies deform into edge states near $z_0=\pm 100$ nm. The green dot in Fig. \ref{Fig:arcs}(h) and the green curve in Fig. \ref{Fig:arcs}(i) show that the edge state near $z_0= 100$ nm mainly distributes near the top surface at $y=50$ nm. By contrast, the edge state near $z_0= -100$ nm mainly distributes near the bottom surface (orange dot and curve). This unique 3D distribution of the edge states of the Fermi arcs can be probed by a combined measurement of in-plane transport and STM. Different from topological insulators \cite{Xu14np,Yoshimi15nc}, the Fermi-arc quantum Hall effect requires the collaboration of the two surfaces.
Note that a 100-nm slab is still a 3D object. Therefore, the quantum Hall effect at Weyl nodes $E_F=E_w$ and the Fermi energy dependence can serve as transport signatures of the Fermi arcs.
The above picture for the Fermi-arc quantum Hall effect can work for Weyl semimetals \cite{Weng15prx,Huang15nc,Lv15prx,Xu15sci-TaAs,HuangXC15prx,Yang15np,Shekhar15np,ZhangCL16nc,Ruan16nc}.

{\color{red}\emph{Topological Dirac semimetals}} - Because of time-reversal symmetry, a single surface of the Dirac semimetal, such as Cd$_3$As$_2$ and Na$_3$Bi, can support a complete Fermi loop required by the quantum Hall effect. The same-surface Fermi arc loop is not that robust and may get deformed \cite{Kargarian16pnas}, and thus may show different characteristics (such as positions and widths of the Hall plateaus) compared to the two-surface Fermi arc loop. The spectrum and Fermi-arc Hall effect in Dirac semimetals can be studied (Secs. S5 and S6 of \cite{Supp}) by using the Hamiltonian \cite{Wang12prb,Wang13prb,Jeon14nmat}
\begin{eqnarray}\label{Eq:H-Wang}
  H&=&\varepsilon_0(\bm k)+
                      \begin{bmatrix}
                        M(\bm k) & Ak_+ & 0 & 0 \\
                        Ak_- & -M(\bm k) & 0 & 0 \\
                        0 & 0 & M(\bm k) & -Ak_- \\
                        0 & 0 & -Ak_+ & -M(\bm k) \\
                      \end{bmatrix}
                      \nonumber\\
&&  +\frac{\mu_B}{2}({\bm \sigma}\cdot\bm B)\otimes
                                                   \begin{bmatrix}
                                                     g_s & 0 \\
                                                     0 & g_p \\
                                                   \end{bmatrix},
\end{eqnarray}
where $g_s$ and $g_p$ are the $g$ factors for the $s$ and $p$ bands \cite{Jeon14nmat}, $k_\pm=k_x\pm ik_y$,
$  \varepsilon_0(\bm k)=C_0+C_1k_z^2+C_2(k_x^2+k_y^2)$, and $  M(\bm k)=M_0+M_1k_z^2+M_2(k_x^2+k_y^2)$. The $x$, $y$, and $z$ axes in the Hamiltonian are defined along the [100], [010], and [001] crystallographic directions, respectively. The samples of Cd$_3$As$_2$ are usually cleaved or grown along [112] or [110] directions, which can be defined as the new $y'$ axis for convenience, as shown in Fig. \ref{Fig:Dirac}(a). For the [112] slab, the parameters \cite{Cano17prbrc} yield that the Fermi arc bands are close to the bulk subbands [Fig. \ref{Fig:Dirac}(c)], implying that the quantum Hall effect may exhibit a fourfold degeneracy. For the [110] slab, the quantum Hall effect may come from pure Fermi arc states [Fig. \ref{Fig:Dirac}(d)]. Figure \ref{Fig:Dirac}(e) shows that for the [110] Cd$_3$As$_2$ slab the odd plateaus are wider than the even plateaus, because the $g$ factor is large. This feature is robust when rotating the magnetic field. The Na$_3$Bi samples cleaved along the [010] direction \cite{Xu15sci} can be used to probe the quantum Hall effect of the Fermi arcs [Figs. \ref{Fig:Dirac}(f) and \ref{Fig:Dirac}(g)]. The $C$ and $M$ terms in Eq. (\ref{Eq:H-Wang}) secure the 2D Fermi arc on the (010) surface.

%A larger Fermi arc loop in Na$_3$Bi leads to a larger magnetic field compared to that in Cd$_3$As$_2$.

We thank helpful discussions with Faxian Xiu, Wang Yao, Hongming Weng, Xi Dai, Dapeng Yu, Yusheng Zhao, Wenqing Zhang, Jiaqing He, and Lang Chen. This work was supported by NBRPC (Grant No. 2015CB921102), Guangdong Innovative and Entrepreneurial Research Team Program (Grant No. 2016ZT06D348), the National Key R \& D Program (Grant No. 2016YFA0301700), the National Natural Science Foundation of China (Grants No. 11534001, No. 11474005, and No. 11574127), and the Science, Technology and Innovation Commission of Shenzhen Municipality (Grant No. ZDSYS20170303165926217).

%\bibliographystyle{apsrev4-1-etal-title}
%\bibliography{refs-transport}

\begin{thebibliography}{70}%
\makeatletter
\providecommand \@ifxundefined [1]{%
 \@ifx{#1\undefined}
}%
\providecommand \@ifnum [1]{%
 \ifnum #1\expandafter \@firstoftwo
 \else \expandafter \@secondoftwo
 \fi
}%
\providecommand \@ifx [1]{%
 \ifx #1\expandafter \@firstoftwo
 \else \expandafter \@secondoftwo
 \fi
}%
\providecommand \natexlab [1]{#1}%
\providecommand \enquote  [1]{``#1''}%
\providecommand \bibnamefont  [1]{#1}%
\providecommand \bibfnamefont [1]{#1}%
\providecommand \citenamefont [1]{#1}%
\providecommand \href@noop [0]{\@secondoftwo}%
\providecommand \href [0]{\begingroup \@sanitize@url \@href}%
\providecommand \@href[1]{\@@startlink{#1}\@@href}%
\providecommand \@@href[1]{\endgroup#1\@@endlink}%
\providecommand \@sanitize@url [0]{\catcode `\\12\catcode `\$12\catcode
  `\&12\catcode `\#12\catcode `\^12\catcode `\_12\catcode `\%12\relax}%
\providecommand \@@startlink[1]{}%
\providecommand \@@endlink[0]{}%
\providecommand \url  [0]{\begingroup\@sanitize@url \@url }%
\providecommand \@url [1]{\endgroup\@href {#1}{\urlprefix }}%
\providecommand \urlprefix  [0]{URL }%
\providecommand \Eprint [0]{\href }%
\providecommand \doibase [0]{http://dx.doi.org/}%
\providecommand \selectlanguage [0]{\@gobble}%
\providecommand \bibinfo  [0]{\@secondoftwo}%
\providecommand \bibfield  [0]{\@secondoftwo}%
\providecommand \translation [1]{[#1]}%
\providecommand \BibitemOpen [0]{}%
\providecommand \bibitemStop [0]{}%
\providecommand \bibitemNoStop [0]{.\EOS\space}%
\providecommand \EOS [0]{\spacefactor3000\relax}%
\providecommand \BibitemShut  [1]{\csname bibitem#1\endcsname}%
\let\auto@bib@innerbib\@empty
%</preamble>
\bibitem [{\citenamefont {Klitzing}\ \emph {et~al.}(1980)\citenamefont
  {Klitzing}, \citenamefont {Dorda},\ and\ \citenamefont
  {Pepper}}]{Klitzing80prl}%
  \BibitemOpen
  \bibfield  {author} {\bibinfo {author} {\bibfnamefont {K.~v.}\ \bibnamefont
  {Klitzing}}, \bibinfo {author} {\bibfnamefont {G.}~\bibnamefont {Dorda}}, \
  and\ \bibinfo {author} {\bibfnamefont {M.}~\bibnamefont {Pepper}},\
  }\bibfield  {title} {\enquote {\bibinfo {title} {New method for high-accuracy
  determination of the fine-structure constant based on quantized hall
  resistance},}\ }\href {\doibase 10.1103/PhysRevLett.45.494} {\bibfield
  {journal} {\bibinfo  {journal} {Phys. Rev. Lett.}\ }\textbf {\bibinfo
  {volume} {45}},\ \bibinfo {pages} {494} (\bibinfo {year} {1980})}\BibitemShut
  {NoStop}%
\bibitem [{\citenamefont {Thouless}\ \emph {et~al.}(1982)\citenamefont
  {Thouless}, \citenamefont {Kohmoto}, \citenamefont {Nightingale},\ and\
  \citenamefont {den Nijs}}]{Thouless82prl}%
  \BibitemOpen
  \bibfield  {author} {\bibinfo {author} {\bibfnamefont {D.~J.}\ \bibnamefont
  {Thouless}}, \bibinfo {author} {\bibfnamefont {M.}~\bibnamefont {Kohmoto}},
  \bibinfo {author} {\bibfnamefont {M.~P.}\ \bibnamefont {Nightingale}}, \ and\
  \bibinfo {author} {\bibfnamefont {M.}~\bibnamefont {den Nijs}},\ }\bibfield
  {title} {\enquote {\bibinfo {title} {Quantized {Hall} conductance in a
  two-dimensional periodic potential},}\ }\href {\doibase
  10.1103/PhysRevLett.49.405} {\bibfield  {journal} {\bibinfo  {journal} {Phys.
  Rev. Lett.}\ }\textbf {\bibinfo {volume} {49}},\ \bibinfo {pages} {405}
  (\bibinfo {year} {1982})}\BibitemShut {NoStop}%
\bibitem [{\citenamefont {Novoselov}\ \emph {et~al.}(2005)\citenamefont
  {Novoselov}, \citenamefont {Geim}, \citenamefont {Morozov}, \citenamefont
  {Jiang}, \citenamefont {Katsnelson}, \citenamefont {Grigorieva},
  \citenamefont {Dubonos},\ and\ \citenamefont {Firsov}}]{Novoselov05nat}%
  \BibitemOpen
  \bibfield  {author} {\bibinfo {author} {\bibfnamefont {K.~S.}\ \bibnamefont
  {Novoselov}}, \bibinfo {author} {\bibfnamefont {A.~K.}\ \bibnamefont {Geim}},
  \bibinfo {author} {\bibfnamefont {S.~V.}\ \bibnamefont {Morozov}}, \bibinfo
  {author} {\bibfnamefont {D.}~\bibnamefont {Jiang}}, \bibinfo {author}
  {\bibfnamefont {M.~I.}\ \bibnamefont {Katsnelson}}, \bibinfo {author}
  {\bibfnamefont {I.~V.}\ \bibnamefont {Grigorieva}}, \bibinfo {author}
  {\bibfnamefont {S.~V.}\ \bibnamefont {Dubonos}}, \ and\ \bibinfo {author}
  {\bibfnamefont {A.~A.}\ \bibnamefont {Firsov}},\ }\bibfield  {title}
  {\enquote {\bibinfo {title} {Two-dimensional gas of massless {Dirac} fermions
  in graphene},}\ }\href {\doibase 10.1038/nature04233} {\bibfield  {journal}
  {\bibinfo  {journal} {Nature}\ }\textbf {\bibinfo {volume} {438}},\ \bibinfo
  {pages} {197} (\bibinfo {year} {2005})}\BibitemShut {NoStop}%
\bibitem [{\citenamefont {Zhang}\ \emph {et~al.}(2005)\citenamefont {Zhang},
  \citenamefont {Tan}, \citenamefont {Stormer},\ and\ \citenamefont
  {Kim}}]{ZhangYB05nat}%
  \BibitemOpen
  \bibfield  {author} {\bibinfo {author} {\bibfnamefont {Y.}~\bibnamefont
  {Zhang}}, \bibinfo {author} {\bibfnamefont {Y.-W.}\ \bibnamefont {Tan}},
  \bibinfo {author} {\bibfnamefont {H.~L.}\ \bibnamefont {Stormer}}, \ and\
  \bibinfo {author} {\bibfnamefont {P.}~\bibnamefont {Kim}},\ }\bibfield
  {title} {\enquote {\bibinfo {title} {Experimental observation of the quantum
  {Hall} effect and {Berry's} phase in graphene},}\ }\href {\doibase
  10.1038/nature04235} {\bibfield  {journal} {\bibinfo  {journal} {Nature}\
  }\textbf {\bibinfo {volume} {438}},\ \bibinfo {pages} {201} (\bibinfo {year}
  {2005})}\BibitemShut {NoStop}%
\bibitem [{\citenamefont {Xu}\ \emph {et~al.}(2014)\citenamefont {Xu},
  \citenamefont {Miotkowski}, \citenamefont {Liu}, \citenamefont {Tian},
  \citenamefont {Nam}, \citenamefont {Alidoust}, \citenamefont {Hu},
  \citenamefont {Shih}, \citenamefont {Hasan},\ and\ \citenamefont
  {Chen}}]{Xu14np}%
  \BibitemOpen
  \bibfield  {author} {\bibinfo {author} {\bibfnamefont {Y.}~\bibnamefont
  {Xu}}, \bibinfo {author} {\bibfnamefont {I.}~\bibnamefont {Miotkowski}},
  \bibinfo {author} {\bibfnamefont {C.}~\bibnamefont {Liu}}, \bibinfo {author}
  {\bibfnamefont {J.}~\bibnamefont {Tian}}, \bibinfo {author} {\bibfnamefont
  {H.}~\bibnamefont {Nam}}, \bibinfo {author} {\bibfnamefont {N.}~\bibnamefont
  {Alidoust}}, \bibinfo {author} {\bibfnamefont {J.}~\bibnamefont {Hu}},
  \bibinfo {author} {\bibfnamefont {C.-K.}\ \bibnamefont {Shih}}, \bibinfo
  {author} {\bibfnamefont {M.~Z.}\ \bibnamefont {Hasan}}, \ and\ \bibinfo
  {author} {\bibfnamefont {Y.~P.}\ \bibnamefont {Chen}},\ }\bibfield  {title}
  {\enquote {\bibinfo {title} {Observation of topological surface state quantum
  {Hall} effect in an intrinsic three-dimensional topological insulator},}\
  }\href {\doibase 10.1038/nphys3140} {\bibfield  {journal} {\bibinfo
  {journal} {Nature Phys.}\ }\textbf {\bibinfo {volume} {10}},\ \bibinfo
  {pages} {956} (\bibinfo {year} {2014})}\BibitemShut {NoStop}%
\bibitem [{\citenamefont {Yoshimi}\ \emph {et~al.}(2015)\citenamefont
  {Yoshimi}, \citenamefont {Yasuda}, \citenamefont {Tsukazaki}, \citenamefont
  {Takahashi}, \citenamefont {Nagaosa}, \citenamefont {Kawasaki},\ and\
  \citenamefont {Tokura}}]{Yoshimi15nc}%
  \BibitemOpen
  \bibfield  {author} {\bibinfo {author} {\bibfnamefont {R.}~\bibnamefont
  {Yoshimi}}, \bibinfo {author} {\bibfnamefont {K.}~\bibnamefont {Yasuda}},
  \bibinfo {author} {\bibfnamefont {A.}~\bibnamefont {Tsukazaki}}, \bibinfo
  {author} {\bibfnamefont {K.~S.}\ \bibnamefont {Takahashi}}, \bibinfo {author}
  {\bibfnamefont {N.}~\bibnamefont {Nagaosa}}, \bibinfo {author} {\bibfnamefont
  {M.}~\bibnamefont {Kawasaki}}, \ and\ \bibinfo {author} {\bibfnamefont
  {Y.}~\bibnamefont {Tokura}},\ }\bibfield  {title} {\enquote {\bibinfo {title}
  {Quantum {Hall} states stabilized in semi-magnetic bilayers of topological
  insulators},}\ }\href {http://www.nature.com/articles/ncomms9530} {\bibfield
  {journal} {\bibinfo  {journal} {Nature Commun.}\ }\textbf {\bibinfo {volume}
  {6}},\ \bibinfo {pages} {8530} (\bibinfo {year} {2015})}\BibitemShut
  {NoStop}%
\bibitem [{\citenamefont {Wan}\ \emph {et~al.}(2011)\citenamefont {Wan},
  \citenamefont {Turner}, \citenamefont {Vishwanath},\ and\ \citenamefont
  {Savrasov}}]{Wan11prb}%
  \BibitemOpen
  \bibfield  {author} {\bibinfo {author} {\bibfnamefont {X.}~\bibnamefont
  {Wan}}, \bibinfo {author} {\bibfnamefont {A.~M.}\ \bibnamefont {Turner}},
  \bibinfo {author} {\bibfnamefont {A.}~\bibnamefont {Vishwanath}}, \ and\
  \bibinfo {author} {\bibfnamefont {S.~Y.}\ \bibnamefont {Savrasov}},\
  }\bibfield  {title} {\enquote {\bibinfo {title} {Topological semimetal and
  {Fermi}-arc surface states in the electronic structure of pyrochlore
  iridates},}\ }\href {\doibase 10.1103/PhysRevB.83.205101} {\bibfield
  {journal} {\bibinfo  {journal} {Phys. Rev. B}\ }\textbf {\bibinfo {volume}
  {83}},\ \bibinfo {pages} {205101} (\bibinfo {year} {2011})}\BibitemShut
  {NoStop}%
\bibitem [{\citenamefont {Yang}\ \emph {et~al.}(2011)\citenamefont {Yang},
  \citenamefont {Lu},\ and\ \citenamefont {Ran}}]{Yang11prb}%
  \BibitemOpen
  \bibfield  {author} {\bibinfo {author} {\bibfnamefont {K.~Y.}\ \bibnamefont
  {Yang}}, \bibinfo {author} {\bibfnamefont {Y.~M.}\ \bibnamefont {Lu}}, \ and\
  \bibinfo {author} {\bibfnamefont {Y.}~\bibnamefont {Ran}},\ }\bibfield
  {title} {\enquote {\bibinfo {title} {Quantum {Hall} effects in a {Weyl}
  semimetal: Possible application in pyrochlore iridates},}\ }\href {\doibase
  10.1103/PhysRevB.84.075129} {\bibfield  {journal} {\bibinfo  {journal} {Phys.
  Rev. B}\ }\textbf {\bibinfo {volume} {84}},\ \bibinfo {pages} {075129}
  (\bibinfo {year} {2011})}\BibitemShut {NoStop}%
\bibitem [{\citenamefont {Burkov}\ and\ \citenamefont
  {Balents}(2011)}]{Burkov11prl}%
  \BibitemOpen
  \bibfield  {author} {\bibinfo {author} {\bibfnamefont {A.~A.}\ \bibnamefont
  {Burkov}}\ and\ \bibinfo {author} {\bibfnamefont {L.}~\bibnamefont
  {Balents}},\ }\bibfield  {title} {\enquote {\bibinfo {title} {{Weyl}
  semimetal in a topological insulator multilayer},}\ }\href {\doibase
  10.1103/PhysRevLett.107.127205} {\bibfield  {journal} {\bibinfo  {journal}
  {Phys. Rev. Lett.}\ }\textbf {\bibinfo {volume} {107}},\ \bibinfo {pages}
  {127205} (\bibinfo {year} {2011})}\BibitemShut {NoStop}%
\bibitem [{\citenamefont {Xu}\ \emph {et~al.}(2011)\citenamefont {Xu},
  \citenamefont {Weng}, \citenamefont {Wang}, \citenamefont {Dai},\ and\
  \citenamefont {Fang}}]{Xu11prl}%
  \BibitemOpen
  \bibfield  {author} {\bibinfo {author} {\bibfnamefont {G.}~\bibnamefont
  {Xu}}, \bibinfo {author} {\bibfnamefont {H.~M.}\ \bibnamefont {Weng}},
  \bibinfo {author} {\bibfnamefont {Z.~J.}\ \bibnamefont {Wang}}, \bibinfo
  {author} {\bibfnamefont {X.}~\bibnamefont {Dai}}, \ and\ \bibinfo {author}
  {\bibfnamefont {Z.}~\bibnamefont {Fang}},\ }\bibfield  {title} {\enquote
  {\bibinfo {title} {Chern semimetal and the quantized anomalous {Hall} effect
  in $\text{HgCr}_2\text{Se}_4$},}\ }\href {\doibase
  10.1103/PhysRevLett.107.186806} {\bibfield  {journal} {\bibinfo  {journal}
  {Phys. Rev. Lett.}\ }\textbf {\bibinfo {volume} {107}},\ \bibinfo {pages}
  {186806} (\bibinfo {year} {2011})}\BibitemShut {NoStop}%
\bibitem [{\citenamefont {Delplace}\ \emph {et~al.}(2012)\citenamefont
  {Delplace}, \citenamefont {Li},\ and\ \citenamefont
  {Carpentier}}]{Delplace12epl}%
  \BibitemOpen
  \bibfield  {author} {\bibinfo {author} {\bibfnamefont {P.}~\bibnamefont
  {Delplace}}, \bibinfo {author} {\bibfnamefont {J.}~\bibnamefont {Li}}, \ and\
  \bibinfo {author} {\bibfnamefont {D.}~\bibnamefont {Carpentier}},\ }\bibfield
   {title} {\enquote {\bibinfo {title} {Topological {Weyl} semi-metal from a
  lattice model},}\ }\href {http://stacks.iop.org/0295-5075/97/i=6/a=67004}
  {\bibfield  {journal} {\bibinfo  {journal} {EPL}\ }\textbf {\bibinfo {volume}
  {97}},\ \bibinfo {pages} {67004} (\bibinfo {year} {2012})}\BibitemShut
  {NoStop}%
\bibitem [{\citenamefont {Jiang}(2012)}]{Jiang12pra}%
  \BibitemOpen
  \bibfield  {author} {\bibinfo {author} {\bibfnamefont {J.-H.}\ \bibnamefont
  {Jiang}},\ }\bibfield  {title} {\enquote {\bibinfo {title} {Tunable
  topological {Weyl} semimetal from simple-cubic lattices with staggered
  fluxes},}\ }\href {\doibase 10.1103/PhysRevA.85.033640} {\bibfield  {journal}
  {\bibinfo  {journal} {Phys. Rev. A}\ }\textbf {\bibinfo {volume} {85}},\
  \bibinfo {pages} {033640} (\bibinfo {year} {2012})}\BibitemShut {NoStop}%
\bibitem [{\citenamefont {Young}\ \emph {et~al.}(2012)\citenamefont {Young},
  \citenamefont {Zaheer}, \citenamefont {Teo}, \citenamefont {Kane},
  \citenamefont {Mele},\ and\ \citenamefont {Rappe}}]{Young12prl}%
  \BibitemOpen
  \bibfield  {author} {\bibinfo {author} {\bibfnamefont {S.~M.}\ \bibnamefont
  {Young}}, \bibinfo {author} {\bibfnamefont {S.}~\bibnamefont {Zaheer}},
  \bibinfo {author} {\bibfnamefont {J.~C.~Y.}\ \bibnamefont {Teo}}, \bibinfo
  {author} {\bibfnamefont {C.~L.}\ \bibnamefont {Kane}}, \bibinfo {author}
  {\bibfnamefont {E.~J.}\ \bibnamefont {Mele}}, \ and\ \bibinfo {author}
  {\bibfnamefont {A.~M.}\ \bibnamefont {Rappe}},\ }\bibfield  {title} {\enquote
  {\bibinfo {title} {{Dirac} semimetal in three dimensions},}\ }\href {\doibase
  10.1103/PhysRevLett.108.140405} {\bibfield  {journal} {\bibinfo  {journal}
  {Phys. Rev. Lett.}\ }\textbf {\bibinfo {volume} {108}},\ \bibinfo {pages}
  {140405} (\bibinfo {year} {2012})}\BibitemShut {NoStop}%
\bibitem [{\citenamefont {Wang}\ \emph {et~al.}(2012)\citenamefont {Wang},
  \citenamefont {Sun}, \citenamefont {Chen}, \citenamefont {Franchini},
  \citenamefont {Xu}, \citenamefont {Weng}, \citenamefont {Dai},\ and\
  \citenamefont {Fang}}]{Wang12prb}%
  \BibitemOpen
  \bibfield  {author} {\bibinfo {author} {\bibfnamefont {Z.}~\bibnamefont
  {Wang}}, \bibinfo {author} {\bibfnamefont {Y.}~\bibnamefont {Sun}}, \bibinfo
  {author} {\bibfnamefont {X.~Q.}\ \bibnamefont {Chen}}, \bibinfo {author}
  {\bibfnamefont {C.}~\bibnamefont {Franchini}}, \bibinfo {author}
  {\bibfnamefont {G.}~\bibnamefont {Xu}}, \bibinfo {author} {\bibfnamefont
  {H.}~\bibnamefont {Weng}}, \bibinfo {author} {\bibfnamefont {X.}~\bibnamefont
  {Dai}}, \ and\ \bibinfo {author} {\bibfnamefont {Z.}~\bibnamefont {Fang}},\
  }\bibfield  {title} {\enquote {\bibinfo {title} {{Dirac} semimetal and
  topological phase transitions in $\text{A}_3\text{Bi}$ ($\text{A}=\text{Na}$,
  $\text{K}$, $\text{Rb}$)},}\ }\href {\doibase 10.1103/PhysRevB.85.195320}
  {\bibfield  {journal} {\bibinfo  {journal} {Phys. Rev. B}\ }\textbf {\bibinfo
  {volume} {85}},\ \bibinfo {pages} {195320} (\bibinfo {year}
  {2012})}\BibitemShut {NoStop}%
\bibitem [{\citenamefont {Singh}\ \emph {et~al.}(2012)\citenamefont {Singh},
  \citenamefont {Sharma}, \citenamefont {Lin}, \citenamefont {Hasan},
  \citenamefont {Prasad},\ and\ \citenamefont {Bansil}}]{Singh12prb}%
  \BibitemOpen
  \bibfield  {author} {\bibinfo {author} {\bibfnamefont {B.}~\bibnamefont
  {Singh}}, \bibinfo {author} {\bibfnamefont {A.}~\bibnamefont {Sharma}},
  \bibinfo {author} {\bibfnamefont {H.}~\bibnamefont {Lin}}, \bibinfo {author}
  {\bibfnamefont {M.~Z.}\ \bibnamefont {Hasan}}, \bibinfo {author}
  {\bibfnamefont {R.}~\bibnamefont {Prasad}}, \ and\ \bibinfo {author}
  {\bibfnamefont {A.}~\bibnamefont {Bansil}},\ }\bibfield  {title} {\enquote
  {\bibinfo {title} {Topological electronic structure and {Weyl} semimetal in
  the $\text{TlBiSe}_2$ class of semiconductors},}\ }\href {\doibase
  10.1103/PhysRevB.86.115208} {\bibfield  {journal} {\bibinfo  {journal} {Phys.
  Rev. B}\ }\textbf {\bibinfo {volume} {86}},\ \bibinfo {pages} {115208}
  (\bibinfo {year} {2012})}\BibitemShut {NoStop}%
\bibitem [{\citenamefont {Wang}\ \emph {et~al.}(2013)\citenamefont {Wang},
  \citenamefont {Weng}, \citenamefont {Wu}, \citenamefont {Dai},\ and\
  \citenamefont {Fang}}]{Wang13prb}%
  \BibitemOpen
  \bibfield  {author} {\bibinfo {author} {\bibfnamefont {Z.}~\bibnamefont
  {Wang}}, \bibinfo {author} {\bibfnamefont {H.}~\bibnamefont {Weng}}, \bibinfo
  {author} {\bibfnamefont {Q.}~\bibnamefont {Wu}}, \bibinfo {author}
  {\bibfnamefont {X.}~\bibnamefont {Dai}}, \ and\ \bibinfo {author}
  {\bibfnamefont {Z.}~\bibnamefont {Fang}},\ }\bibfield  {title} {\enquote
  {\bibinfo {title} {Three-dimensional {Dirac} semimetal and quantum transport
  in $\text{Cd}_3\text{As}_2$},}\ }\href {\doibase 10.1103/PhysRevB.88.125427}
  {\bibfield  {journal} {\bibinfo  {journal} {Phys. Rev. B}\ }\textbf {\bibinfo
  {volume} {88}},\ \bibinfo {pages} {125427} (\bibinfo {year}
  {2013})}\BibitemShut {NoStop}%
\bibitem [{\citenamefont {Liu}\ and\ \citenamefont
  {Vanderbilt}(2014)}]{LiuJP14prb}%
  \BibitemOpen
  \bibfield  {author} {\bibinfo {author} {\bibfnamefont {J.}~\bibnamefont
  {Liu}}\ and\ \bibinfo {author} {\bibfnamefont {D.}~\bibnamefont
  {Vanderbilt}},\ }\bibfield  {title} {\enquote {\bibinfo {title} {Weyl
  semimetals from noncentrosymmetric topological insulators},}\ }\href
  {\doibase 10.1103/PhysRevB.90.155316} {\bibfield  {journal} {\bibinfo
  {journal} {Phys. Rev. B}\ }\textbf {\bibinfo {volume} {90}},\ \bibinfo
  {pages} {155316} (\bibinfo {year} {2014})}\BibitemShut {NoStop}%
\bibitem [{\citenamefont {Bulmash}\ \emph {et~al.}(2014)\citenamefont
  {Bulmash}, \citenamefont {Liu},\ and\ \citenamefont {Qi}}]{Bulmash14prb}%
  \BibitemOpen
  \bibfield  {author} {\bibinfo {author} {\bibfnamefont {D.}~\bibnamefont
  {Bulmash}}, \bibinfo {author} {\bibfnamefont {C.-X.}\ \bibnamefont {Liu}}, \
  and\ \bibinfo {author} {\bibfnamefont {X.-L.}\ \bibnamefont {Qi}},\
  }\bibfield  {title} {\enquote {\bibinfo {title} {Prediction of a {Weyl}
  semimetal in $\text{HgCdMnTe}$},}\ }\href {\doibase
  10.1103/PhysRevB.89.081106} {\bibfield  {journal} {\bibinfo  {journal} {Phys.
  Rev. B}\ }\textbf {\bibinfo {volume} {89}},\ \bibinfo {pages} {081106}
  (\bibinfo {year} {2014})}\BibitemShut {NoStop}%
\bibitem [{\citenamefont {Brahlek}\ \emph {et~al.}(2012)\citenamefont
  {Brahlek}, \citenamefont {Bansal}, \citenamefont {Koirala}, \citenamefont
  {Xu}, \citenamefont {Neupane}, \citenamefont {Liu}, \citenamefont {Hasan},\
  and\ \citenamefont {Oh}}]{Brahlek12prl}%
  \BibitemOpen
  \bibfield  {author} {\bibinfo {author} {\bibfnamefont {M.}~\bibnamefont
  {Brahlek}}, \bibinfo {author} {\bibfnamefont {N.}~\bibnamefont {Bansal}},
  \bibinfo {author} {\bibfnamefont {N.}~\bibnamefont {Koirala}}, \bibinfo
  {author} {\bibfnamefont {S.~Y.}\ \bibnamefont {Xu}}, \bibinfo {author}
  {\bibfnamefont {M.}~\bibnamefont {Neupane}}, \bibinfo {author} {\bibfnamefont
  {C.}~\bibnamefont {Liu}}, \bibinfo {author} {\bibfnamefont {M.~Z.}\
  \bibnamefont {Hasan}}, \ and\ \bibinfo {author} {\bibfnamefont
  {S.}~\bibnamefont {Oh}},\ }\bibfield  {title} {\enquote {\bibinfo {title}
  {Topological-metal to band-insulator transition in
  $(\text{Bi}_{1-x}\text{In}_{x}{)}_{2}\text{Se}_{3}$ thin films},}\ }\href
  {\doibase 10.1103/PhysRevLett.109.186403} {\bibfield  {journal} {\bibinfo
  {journal} {Phys. Rev. Lett.}\ }\textbf {\bibinfo {volume} {109}},\ \bibinfo
  {pages} {186403} (\bibinfo {year} {2012})}\BibitemShut {NoStop}%
\bibitem [{\citenamefont {Wu}\ \emph {et~al.}(2013)\citenamefont {Wu},
  \citenamefont {Brahlek}, \citenamefont {{Valdes Aquilar}}, \citenamefont
  {Stier}, \citenamefont {Morris}, \citenamefont {Lubashevsky}, \citenamefont
  {Bilbro}, \citenamefont {Bansal}, \citenamefont {Oh},\ and\ \citenamefont
  {Armitage}}]{Wu13natphys}%
  \BibitemOpen
  \bibfield  {author} {\bibinfo {author} {\bibfnamefont {L.}~\bibnamefont
  {Wu}}, \bibinfo {author} {\bibfnamefont {M.}~\bibnamefont {Brahlek}},
  \bibinfo {author} {\bibfnamefont {R.}~\bibnamefont {{Valdes Aquilar}}},
  \bibinfo {author} {\bibfnamefont {A.~V.}\ \bibnamefont {Stier}}, \bibinfo
  {author} {\bibfnamefont {C.~M.}\ \bibnamefont {Morris}}, \bibinfo {author}
  {\bibfnamefont {Y.}~\bibnamefont {Lubashevsky}}, \bibinfo {author}
  {\bibfnamefont {L.~S.}\ \bibnamefont {Bilbro}}, \bibinfo {author}
  {\bibfnamefont {N.}~\bibnamefont {Bansal}}, \bibinfo {author} {\bibfnamefont
  {S.}~\bibnamefont {Oh}}, \ and\ \bibinfo {author} {\bibfnamefont {N.~P.}\
  \bibnamefont {Armitage}},\ }\bibfield  {title} {\enquote {\bibinfo {title} {A
  sudden collapse in the transport lifetime across the topological phase
  transition in $(\text{Bi}_{1-x}\text{In}_x)_2\text{Se}_3$},}\ }\href
  {http://dx.doi.org/10.1038/nphys2647} {\bibfield  {journal} {\bibinfo
  {journal} {Nature Phys.}\ }\textbf {\bibinfo {volume} {9}},\ \bibinfo {pages}
  {410} (\bibinfo {year} {2013})}\BibitemShut {NoStop}%
\bibitem [{\citenamefont {Liu}\ \emph {et~al.}(2014{\natexlab{a}})\citenamefont
  {Liu}, \citenamefont {Zhou}, \citenamefont {Zhang}, \citenamefont {Wang},
  \citenamefont {Weng}, \citenamefont {Prabhakaran}, \citenamefont {Mo},
  \citenamefont {Shen}, \citenamefont {Fang}, \citenamefont {Dai},
  \citenamefont {Hussain},\ and\ \citenamefont {Chen}}]{Liu14sci}%
  \BibitemOpen
  \bibfield  {author} {\bibinfo {author} {\bibfnamefont {Z.~K.}\ \bibnamefont
  {Liu}},  \emph {et~al.},\ }\bibfield  {title} {\enquote {\bibinfo {title}
  {Discovery of a three-dimensional topological {Dirac} semimetal,
  $\text{Na}_3\text{Bi}$},}\ }\href {\doibase 10.1126/science.1245085}
  {\bibfield  {journal} {\bibinfo  {journal} {Science}\ }\textbf {\bibinfo
  {volume} {343}},\ \bibinfo {pages} {864} (\bibinfo {year}
  {2014}{\natexlab{a}})}\BibitemShut {NoStop}%
\bibitem [{\citenamefont {Xu}\ \emph {et~al.}(2015{\natexlab{a}})\citenamefont
  {Xu}, \citenamefont {Liu}, \citenamefont {Kushwaha}, \citenamefont {Sankar},
  \citenamefont {Krizan}, \citenamefont {Belopolski}, \citenamefont {Neupane},
  \citenamefont {Bian}, \citenamefont {Alidoust}, \citenamefont {Chang},
  \citenamefont {Jeng}, \citenamefont {Huang}, \citenamefont {Tsai},
  \citenamefont {Lin}, \citenamefont {Shibayev}, \citenamefont {Chou},
  \citenamefont {Cava},\ and\ \citenamefont {Hasan}}]{Xu15sci}%
  \BibitemOpen
  \bibfield  {author} {\bibinfo {author} {\bibfnamefont {S.~Y.}\ \bibnamefont
  {Xu}},  \emph {et~al.},\ }\bibfield  {title} {\enquote {\bibinfo {title}
  {Observation of $\text{Fermi}$ arc surface states in a topological metal},}\
  }\href {\doibase 10.1126/science.1256742} {\bibfield  {journal} {\bibinfo
  {journal} {Science}\ }\textbf {\bibinfo {volume} {347}},\ \bibinfo {pages}
  {294} (\bibinfo {year} {2015}{\natexlab{a}})}\BibitemShut {NoStop}%
\bibitem [{\citenamefont {Liu}\ \emph {et~al.}(2014{\natexlab{b}})\citenamefont
  {Liu}, \citenamefont {Jiang}, \citenamefont {Zhou}, \citenamefont {Wang},
  \citenamefont {Zhang}, \citenamefont {Weng}, \citenamefont {Prabhakaran},
  \citenamefont {Mo}, \citenamefont {Peng}, \citenamefont {Dudin},
  \citenamefont {Kim}, \citenamefont {Hoesch}, \citenamefont {Fang},
  \citenamefont {Dai}, \citenamefont {Shen}, \citenamefont {Feng},
  \citenamefont {Hussain},\ and\ \citenamefont {Chen}}]{Liu14natmat}%
  \BibitemOpen
  \bibfield  {author} {\bibinfo {author} {\bibfnamefont {Z.~K.}\ \bibnamefont
  {Liu}},  \emph {et~al.},\ }\bibfield  {title} {\enquote {\bibinfo {title} {A
  stable three-dimensional topological {Dirac} semimetal
  $\text{Cd}_3\text{As}_2$},}\ }\href {http://dx.doi.org/10.1038/nmat3990}
  {\bibfield  {journal} {\bibinfo  {journal} {Nature Mater.}\ }\textbf
  {\bibinfo {volume} {13}},\ \bibinfo {pages} {677} (\bibinfo {year}
  {2014}{\natexlab{b}})}\BibitemShut {NoStop}%
\bibitem [{\citenamefont {Neupane}\ \emph {et~al.}(2014)\citenamefont
  {Neupane}, \citenamefont {Xu}, \citenamefont {Sankar}, \citenamefont
  {Alidoust}, \citenamefont {Bian}, \citenamefont {Liu}, \citenamefont
  {Belopolski}, \citenamefont {Chang}, \citenamefont {Jeng}, \citenamefont
  {Lin}, \citenamefont {Bansil}, \citenamefont {Chou},\ and\ \citenamefont
  {Hasan}}]{Neupane14nc}%
  \BibitemOpen
  \bibfield  {author} {\bibinfo {author} {\bibfnamefont {M.}~\bibnamefont
  {Neupane}},  \emph {et~al.},\ }\bibfield  {title} {\enquote {\bibinfo {title}
  {Observation of a three-dimensional topological {Dirac} semimetal phase in
  high-mobility $\text{Cd}_3\text{As}_2$},}\ }\href {\doibase
  10.1038/ncomms4786} {\bibfield  {journal} {\bibinfo  {journal} {Nature
  Commun.}\ }\textbf {\bibinfo {volume} {5}},\ \bibinfo {pages} {3786}
  (\bibinfo {year} {2014})}\BibitemShut {NoStop}%
\bibitem [{\citenamefont {Yi}\ \emph {et~al.}(2014)\citenamefont {Yi},
  \citenamefont {Wang}, \citenamefont {Chen}, \citenamefont {Shi},
  \citenamefont {Feng}, \citenamefont {Liang}, \citenamefont {Xie},
  \citenamefont {He}, \citenamefont {He}, \citenamefont {Peng}, \citenamefont
  {Liu}, \citenamefont {Liu}, \citenamefont {Zhao}, \citenamefont {Liu},
  \citenamefont {Dong}, \citenamefont {Zhang}, \citenamefont {Nakatake},
  \citenamefont {Arita}, \citenamefont {Shimada}, \citenamefont {Namatame},
  \citenamefont {Taniguchi}, \citenamefont {Xu}, \citenamefont {Chen},
  \citenamefont {Dai}, \citenamefont {Fang},\ and\ \citenamefont
  {Zhou}}]{Yi14srep}%
  \BibitemOpen
  \bibfield  {author} {\bibinfo {author} {\bibfnamefont {H.}~\bibnamefont
  {Yi}},  \emph {et~al.},\ }\bibfield  {title} {\enquote {\bibinfo {title}
  {Evidence of topological surface state in three-dimensional {Dirac} semimetal
  $\text{Cd}_3\text{As}_2$},}\ }\href
  {http://www.nature.com/srep/2014/140820/srep06106/full/srep06106.html}
  {\bibfield  {journal} {\bibinfo  {journal} {Sci. Rep.}\ }\textbf {\bibinfo
  {volume} {4}},\ \bibinfo {pages} {6106} (\bibinfo {year} {2014})}\BibitemShut
  {NoStop}%
\bibitem [{\citenamefont {Borisenko}\ \emph {et~al.}(2014)\citenamefont
  {Borisenko}, \citenamefont {Gibson}, \citenamefont {Evtushinsky},
  \citenamefont {Zabolotnyy}, \citenamefont {B\"uchner},\ and\ \citenamefont
  {Cava}}]{Borisenko14prl}%
  \BibitemOpen
  \bibfield  {author} {\bibinfo {author} {\bibfnamefont {S.}~\bibnamefont
  {Borisenko}}, \bibinfo {author} {\bibfnamefont {Q.}~\bibnamefont {Gibson}},
  \bibinfo {author} {\bibfnamefont {D.}~\bibnamefont {Evtushinsky}}, \bibinfo
  {author} {\bibfnamefont {V.}~\bibnamefont {Zabolotnyy}}, \bibinfo {author}
  {\bibfnamefont {B.}~\bibnamefont {B\"uchner}}, \ and\ \bibinfo {author}
  {\bibfnamefont {R.~J.}\ \bibnamefont {Cava}},\ }\bibfield  {title} {\enquote
  {\bibinfo {title} {Experimental realization of a three-dimensional {Dirac}
  semimetal},}\ }\href {\doibase 10.1103/PhysRevLett.113.027603} {\bibfield
  {journal} {\bibinfo  {journal} {Phys. Rev. Lett.}\ }\textbf {\bibinfo
  {volume} {113}},\ \bibinfo {pages} {027603} (\bibinfo {year}
  {2014})}\BibitemShut {NoStop}%
\bibitem [{\citenamefont {Weng}\ \emph {et~al.}(2015)\citenamefont {Weng},
  \citenamefont {Fang}, \citenamefont {Fang}, \citenamefont {Bernevig},\ and\
  \citenamefont {Dai}}]{Weng15prx}%
  \BibitemOpen
  \bibfield  {author} {\bibinfo {author} {\bibfnamefont {H.~M.}\ \bibnamefont
  {Weng}}, \bibinfo {author} {\bibfnamefont {C.}~\bibnamefont {Fang}}, \bibinfo
  {author} {\bibfnamefont {Z.}~\bibnamefont {Fang}}, \bibinfo {author}
  {\bibfnamefont {B.~A.}\ \bibnamefont {Bernevig}}, \ and\ \bibinfo {author}
  {\bibfnamefont {X.}~\bibnamefont {Dai}},\ }\bibfield  {title} {\enquote
  {\bibinfo {title} {Weyl semimetal phase in noncentrosymmetric
  transition-metal monophosphides},}\ }\href {\doibase
  10.1103/PhysRevX.5.011029} {\bibfield  {journal} {\bibinfo  {journal} {Phys.
  Rev. X}\ }\textbf {\bibinfo {volume} {5}},\ \bibinfo {pages} {011029}
  (\bibinfo {year} {2015})}\BibitemShut {NoStop}%
\bibitem [{\citenamefont {Huang}\ \emph
  {et~al.}(2015{\natexlab{a}})\citenamefont {Huang}, \citenamefont {Xu},
  \citenamefont {Belopolski}, \citenamefont {Lee}, \citenamefont {Chang},
  \citenamefont {Wang}, \citenamefont {Alidoust}, \citenamefont {Bian},
  \citenamefont {Neupane}, \citenamefont {Zhang}, \citenamefont {Jia},
  \citenamefont {Bansil}, \citenamefont {Lin},\ and\ \citenamefont
  {Hasan}}]{Huang15nc}%
  \BibitemOpen
  \bibfield  {author} {\bibinfo {author} {\bibfnamefont {S.~M.}\ \bibnamefont
  {Huang}},  \emph {et~al.},\ }\bibfield  {title} {\enquote {\bibinfo {title}
  {A {Weyl} fermion semimetal with surface $\text{Fermi}$ arcs in the
  transition metal monopnictide $\text{TaAs}$ class},}\ }\href {\doibase
  10.1038/ncomms8373} {\bibfield  {journal} {\bibinfo  {journal} {Nat.
  Commun.}\ }\textbf {\bibinfo {volume} {6}},\ \bibinfo {pages} {7373}
  (\bibinfo {year} {2015}{\natexlab{a}})}\BibitemShut {NoStop}%
\bibitem [{\citenamefont {Lv}\ \emph {et~al.}(2015)\citenamefont {Lv},
  \citenamefont {Weng}, \citenamefont {Fu}, \citenamefont {Wang}, \citenamefont
  {Miao}, \citenamefont {Ma}, \citenamefont {Richard}, \citenamefont {Huang},
  \citenamefont {Zhao}, \citenamefont {Chen}, \citenamefont {Fang},
  \citenamefont {Dai}, \citenamefont {Qian},\ and\ \citenamefont
  {Ding}}]{Lv15prx}%
  \BibitemOpen
  \bibfield  {author} {\bibinfo {author} {\bibfnamefont {B.~Q.}\ \bibnamefont
  {Lv}},  \emph {et~al.},\ }\bibfield  {title} {\enquote {\bibinfo {title}
  {Experimental discovery of {Weyl} semimetal $\text{TaAs}$},}\ }\href
  {\doibase 10.1103/PhysRevX.5.031013} {\bibfield  {journal} {\bibinfo
  {journal} {Phys. Rev. X}\ }\textbf {\bibinfo {volume} {5}},\ \bibinfo {pages}
  {031013} (\bibinfo {year} {2015})}\BibitemShut {NoStop}%
\bibitem [{\citenamefont {Xu}\ \emph {et~al.}(2015{\natexlab{b}})\citenamefont
  {Xu}, \citenamefont {Belopolski}, \citenamefont {Alidoust}, \citenamefont
  {Neupane}, \citenamefont {Bian}, \citenamefont {Zhang}, \citenamefont
  {Sankar}, \citenamefont {Chang}, \citenamefont {Yuan}, \citenamefont {Lee},
  \citenamefont {Huang}, \citenamefont {Zheng}, \citenamefont {Ma},
  \citenamefont {Sanchez}, \citenamefont {Wang}, \citenamefont {Bansil},
  \citenamefont {Chou}, \citenamefont {Shibayev}, \citenamefont {Lin},
  \citenamefont {Jia},\ and\ \citenamefont {Hasan}}]{Xu15sci-TaAs}%
  \BibitemOpen
  \bibfield  {author} {\bibinfo {author} {\bibfnamefont {S.~Y.}\ \bibnamefont
  {Xu}},  \emph {et~al.},\ }\bibfield  {title} {\enquote {\bibinfo {title}
  {Discovery of a {Weyl} fermion semimetal and topological $\text{Fermi}$
  arcs},}\ }\href {\doibase 10.1126/science.aaa9297} {\bibfield  {journal}
  {\bibinfo  {journal} {Science}\ }\textbf {\bibinfo {volume} {349}},\ \bibinfo
  {pages} {613} (\bibinfo {year} {2015}{\natexlab{b}})}\BibitemShut {NoStop}%
\bibitem [{\citenamefont {Potter}\ \emph {et~al.}(2014)\citenamefont {Potter},
  \citenamefont {Kimchi},\ and\ \citenamefont {Vishwanath}}]{Potter14nc}%
  \BibitemOpen
  \bibfield  {author} {\bibinfo {author} {\bibfnamefont {A.~C.}\ \bibnamefont
  {Potter}}, \bibinfo {author} {\bibfnamefont {I.}~\bibnamefont {Kimchi}}, \
  and\ \bibinfo {author} {\bibfnamefont {A.}~\bibnamefont {Vishwanath}},\
  }\bibfield  {title} {\enquote {\bibinfo {title} {Quantum oscillations from
  surface $\text{Fermi}$ arcs in {Weyl} and {Dirac} semimetals},}\ }\href
  {\doibase 10.1038/ncomms6161} {\bibfield  {journal} {\bibinfo  {journal}
  {Nature Commun.}\ }\textbf {\bibinfo {volume} {5}},\ \bibinfo {pages} {5161}
  (\bibinfo {year} {2014})}\BibitemShut {NoStop}%
\bibitem [{\citenamefont {Moll}\ \emph {et~al.}(2016)\citenamefont {Moll},
  \citenamefont {Nair}, \citenamefont {Helm}, \citenamefont {Potter},
  \citenamefont {Kimchi}, \citenamefont {Vishwanath},\ and\ \citenamefont
  {Analytis}}]{Moll16nat}%
  \BibitemOpen
  \bibfield  {author} {\bibinfo {author} {\bibfnamefont {P.~J.~W.}\
  \bibnamefont {Moll}}, \bibinfo {author} {\bibfnamefont {N.~L.}\ \bibnamefont
  {Nair}}, \bibinfo {author} {\bibfnamefont {T.}~\bibnamefont {Helm}}, \bibinfo
  {author} {\bibfnamefont {A.~C.}\ \bibnamefont {Potter}}, \bibinfo {author}
  {\bibfnamefont {I.}~\bibnamefont {Kimchi}}, \bibinfo {author} {\bibfnamefont
  {A.}~\bibnamefont {Vishwanath}}, \ and\ \bibinfo {author} {\bibfnamefont
  {J.~G.}\ \bibnamefont {Analytis}},\ }\bibfield  {title} {\enquote {\bibinfo
  {title} {Transport evidence for {Fermi-arc-mediated} chirality transfer in
  the {Dirac} semimetal {Cd}$_3${As}$_2$},}\ }\href {\doibase
  10.1038/nature18276} {\bibfield  {journal} {\bibinfo  {journal} {Nature}\
  }\textbf {\bibinfo {volume} {535}},\ \bibinfo {pages} {266} (\bibinfo {year}
  {2016})}\BibitemShut {NoStop}%
\bibitem [{\citenamefont {Hosur}(2012)}]{Hosur12prb}%
  \BibitemOpen
  \bibfield  {author} {\bibinfo {author} {\bibfnamefont {P.}~\bibnamefont
  {Hosur}},\ }\bibfield  {title} {\enquote {\bibinfo {title} {Friedel
  oscillations due to $\text{Fermi}$ arcs in {Weyl} semimetals},}\ }\href
  {\doibase 10.1103/PhysRevB.86.195102} {\bibfield  {journal} {\bibinfo
  {journal} {Phys. Rev. B}\ }\textbf {\bibinfo {volume} {86}},\ \bibinfo
  {pages} {195102} (\bibinfo {year} {2012})}\BibitemShut {NoStop}%
\bibitem [{\citenamefont {Baum}\ \emph {et~al.}(2015)\citenamefont {Baum},
  \citenamefont {Berg}, \citenamefont {Parameswaran},\ and\ \citenamefont
  {Stern}}]{Baum15prx}%
  \BibitemOpen
  \bibfield  {author} {\bibinfo {author} {\bibfnamefont {Y.}~\bibnamefont
  {Baum}}, \bibinfo {author} {\bibfnamefont {E.}~\bibnamefont {Berg}}, \bibinfo
  {author} {\bibfnamefont {S.~A.}\ \bibnamefont {Parameswaran}}, \ and\
  \bibinfo {author} {\bibfnamefont {A.}~\bibnamefont {Stern}},\ }\bibfield
  {title} {\enquote {\bibinfo {title} {Current at a distance and resonant
  transparency in {Weyl} semimetals},}\ }\href {\doibase
  10.1103/PhysRevX.5.041046} {\bibfield  {journal} {\bibinfo  {journal} {Phys.
  Rev. X}\ }\textbf {\bibinfo {volume} {5}},\ \bibinfo {pages} {041046}
  (\bibinfo {year} {2015})}\BibitemShut {NoStop}%
\bibitem [{\citenamefont {Gorbar}\ \emph {et~al.}(2016)\citenamefont {Gorbar},
  \citenamefont {Miransky}, \citenamefont {Shovkovy},\ and\ \citenamefont
  {Sukhachov}}]{Gorbar16prb}%
  \BibitemOpen
  \bibfield  {author} {\bibinfo {author} {\bibfnamefont {E.~V.}\ \bibnamefont
  {Gorbar}}, \bibinfo {author} {\bibfnamefont {V.~A.}\ \bibnamefont
  {Miransky}}, \bibinfo {author} {\bibfnamefont {I.~A.}\ \bibnamefont
  {Shovkovy}}, \ and\ \bibinfo {author} {\bibfnamefont {P.~O.}\ \bibnamefont
  {Sukhachov}},\ }\bibfield  {title} {\enquote {\bibinfo {title} {Origin of
  dissipative {Fermi} arc transport in {Weyl} semimetals},}\ }\href {\doibase
  10.1103/PhysRevB.93.235127} {\bibfield  {journal} {\bibinfo  {journal} {Phys.
  Rev. B}\ }\textbf {\bibinfo {volume} {93}},\ \bibinfo {pages} {235127}
  (\bibinfo {year} {2016})}\BibitemShut {NoStop}%
\bibitem [{\citenamefont {Ominato}\ and\ \citenamefont
  {Koshino}(2016)}]{Ominato16prb}%
  \BibitemOpen
  \bibfield  {author} {\bibinfo {author} {\bibfnamefont {Y.}~\bibnamefont
  {Ominato}}\ and\ \bibinfo {author} {\bibfnamefont {M.}~\bibnamefont
  {Koshino}},\ }\bibfield  {title} {\enquote {\bibinfo {title}
  {Magnetotransport in {Weyl} semimetals in the quantum limit: Role of
  topological surface states},}\ }\href {\doibase 10.1103/PhysRevB.93.245304}
  {\bibfield  {journal} {\bibinfo  {journal} {Phys. Rev. B}\ }\textbf {\bibinfo
  {volume} {93}},\ \bibinfo {pages} {245304} (\bibinfo {year}
  {2016})}\BibitemShut {NoStop}%
\bibitem [{\citenamefont {McCormick}\ \emph {et~al.}(2017)\citenamefont
  {McCormick}, \citenamefont {Watzman}, \citenamefont {Heremans},\ and\
  \citenamefont {Trivedi}}]{McCormick17arXiv}%
  \BibitemOpen
  \bibfield  {author} {\bibinfo {author} {\bibfnamefont {T.~M.}\ \bibnamefont
  {McCormick}}, \bibinfo {author} {\bibfnamefont {S.~J.}\ \bibnamefont
  {Watzman}}, \bibinfo {author} {\bibfnamefont {J.~P.}\ \bibnamefont
  {Heremans}}, \ and\ \bibinfo {author} {\bibfnamefont {N.}~\bibnamefont
  {Trivedi}},\ }\bibfield  {title} {\enquote {\bibinfo {title} {Fermi arc
  mediated entropy transport in topological semimetals},}\ }\href
  {https://arxiv.org/abs/1703.04606} {\bibfield  {journal} {\bibinfo  {journal}
  {arXiv:1703.04606}\ } (\bibinfo {year} {2017})}\BibitemShut {NoStop}%
\bibitem [{\citenamefont {Zhang}\ \emph
  {et~al.}(2016{\natexlab{a}})\citenamefont {Zhang}, \citenamefont {Lu},\ and\
  \citenamefont {Shen}}]{ZhangSB16njp}%
  \BibitemOpen
  \bibfield  {author} {\bibinfo {author} {\bibfnamefont {S.-B.}\ \bibnamefont
  {Zhang}}, \bibinfo {author} {\bibfnamefont {H.-Z.}\ \bibnamefont {Lu}}, \
  and\ \bibinfo {author} {\bibfnamefont {S.-Q.}\ \bibnamefont {Shen}},\
  }\bibfield  {title} {\enquote {\bibinfo {title} {Linear magnetoconductivity
  in an intrinsic topological {Weyl} semimetal},}\ }\href
  {http://stacks.iop.org/1367-2630/18/i=5/a=053039} {\bibfield  {journal}
  {\bibinfo  {journal} {New J. Phys.}\ }\textbf {\bibinfo {volume} {18}},\
  \bibinfo {pages} {053039} (\bibinfo {year} {2016}{\natexlab{a}})}\BibitemShut
  {NoStop}%
\bibitem [{\citenamefont {Ruan}\ \emph {et~al.}(2016)\citenamefont {Ruan},
  \citenamefont {Jian}, \citenamefont {Yao}, \citenamefont {Zhang},
  \citenamefont {Zhang},\ and\ \citenamefont {Xing}}]{Ruan16nc}%
  \BibitemOpen
  \bibfield  {author} {\bibinfo {author} {\bibfnamefont {J.}~\bibnamefont
  {Ruan}}, \bibinfo {author} {\bibfnamefont {S.-K.}\ \bibnamefont {Jian}},
  \bibinfo {author} {\bibfnamefont {H.}~\bibnamefont {Yao}}, \bibinfo {author}
  {\bibfnamefont {H.}~\bibnamefont {Zhang}}, \bibinfo {author} {\bibfnamefont
  {S.-C.}\ \bibnamefont {Zhang}}, \ and\ \bibinfo {author} {\bibfnamefont
  {D.}~\bibnamefont {Xing}},\ }\bibfield  {title} {\enquote {\bibinfo {title}
  {Symmetry-protected ideal {Weyl} semimetal in {HgTe}-class materials},}\
  }\href {http://dx.doi.org/10.1038/ncomms11136} {\bibfield  {journal}
  {\bibinfo  {journal} {Nature Commun.}\ }\textbf {\bibinfo {volume} {7}},\
  \bibinfo {pages} {11136} (\bibinfo {year} {2016})}\BibitemShut {NoStop}%
\bibitem [{\citenamefont {Huang}\ \emph
  {et~al.}(2015{\natexlab{b}})\citenamefont {Huang}, \citenamefont {Zhao},
  \citenamefont {Long}, \citenamefont {Wang}, \citenamefont {Chen},
  \citenamefont {Yang}, \citenamefont {Liang}, \citenamefont {Xue},
  \citenamefont {Weng}, \citenamefont {Fang}, \citenamefont {Dai},\ and\
  \citenamefont {Chen}}]{HuangXC15prx}%
  \BibitemOpen
  \bibfield  {author} {\bibinfo {author} {\bibfnamefont {X.~C.}\ \bibnamefont
  {Huang}},  \emph {et~al.},\ }\bibfield  {title} {\enquote {\bibinfo {title}
  {Observation of the chiral-anomaly-induced negative magnetoresistance in
  $\text{3D}$ {Weyl} semimetal $\text{TaAs}$},}\ }\href {\doibase
  10.1103/PhysRevX.5.031023} {\bibfield  {journal} {\bibinfo  {journal} {Phys.
  Rev. X}\ }\textbf {\bibinfo {volume} {5}},\ \bibinfo {pages} {031023}
  (\bibinfo {year} {2015}{\natexlab{b}})}\BibitemShut {NoStop}%
\bibitem [{\citenamefont {Yang}\ \emph {et~al.}(2015)\citenamefont {Yang},
  \citenamefont {Liu}, \citenamefont {Sun}, \citenamefont {Peng}, \citenamefont
  {Yang}, \citenamefont {Zhang}, \citenamefont {Zhou}, \citenamefont {Zhang},
  \citenamefont {Guo}, \citenamefont {Rahn}, \citenamefont {Prabhakaran},
  \citenamefont {Hussain}, \citenamefont {Mo}, \citenamefont {Felser},
  \citenamefont {Yan},\ and\ \citenamefont {Chen}}]{Yang15np}%
  \BibitemOpen
  \bibfield  {author} {\bibinfo {author} {\bibfnamefont {L.~X.}\ \bibnamefont
  {Yang}},  \emph {et~al.},\ }\bibfield  {title} {\enquote {\bibinfo {title}
  {{Weyl} semimetal phase in the non-centrosymmetric compound {TaAs}},}\ }\href
  {http://www.nature.com/nphys/journal/v11/n9/abs/nphys3425.html} {\bibfield
  {journal} {\bibinfo  {journal} {Nature Phys.}\ }\textbf {\bibinfo {volume}
  {11}},\ \bibinfo {pages} {728} (\bibinfo {year} {2015})}\BibitemShut
  {NoStop}%
\bibitem [{\citenamefont {Shekhar}\ \emph {et~al.}(2015)\citenamefont
  {Shekhar}, \citenamefont {Nayak}, \citenamefont {Sun}, \citenamefont
  {Schmidt}, \citenamefont {Nicklas}, \citenamefont {Leermakers}, \citenamefont
  {Zeitler}, \citenamefont {Schnelle}, \citenamefont {Grin}, \citenamefont
  {Felser},\ and\ \citenamefont {Yan}}]{Shekhar15np}%
  \BibitemOpen
  \bibfield  {author} {\bibinfo {author} {\bibfnamefont {C.}~\bibnamefont
  {Shekhar}},  \emph {et~al.},\ }\bibfield  {title} {\enquote {\bibinfo {title}
  {Extremely large magnetoresistance and ultrahigh mobility in the topological
  {Weyl} semimetal $\text{NbP}$},}\ }\href
  {http://www.nature.com/nphys/journal/v11/n8/full/nphys3372.html} {\bibfield
  {journal} {\bibinfo  {journal} {Nature Phys.}\ }\textbf {\bibinfo {volume}
  {11}},\ \bibinfo {pages} {645} (\bibinfo {year} {2015})}\BibitemShut
  {NoStop}%
\bibitem [{\citenamefont {Zhang}\ \emph
  {et~al.}(2016{\natexlab{b}})\citenamefont {Zhang}, \citenamefont {Xu},
  \citenamefont {Belopolski}, \citenamefont {Yuan}, \citenamefont {Lin},
  \citenamefont {Tong}, \citenamefont {Alidoust}, \citenamefont {Lee},
  \citenamefont {Huang}, \citenamefont {Chang}, \citenamefont {Jeng},
  \citenamefont {Lin}, \citenamefont {Neupane}, \citenamefont {Sanchez},
  \citenamefont {Zheng}, \citenamefont {Bian}, \citenamefont {Wang},
  \citenamefont {Zhang}, \citenamefont {Lu}, \citenamefont {Shen},
  \citenamefont {Neupert}, \citenamefont {Hasan},\ and\ \citenamefont
  {Jia}}]{ZhangCL16nc}%
  \BibitemOpen
  \bibfield  {author} {\bibinfo {author} {\bibfnamefont {C.~L.}\ \bibnamefont
  {Zhang}},  \emph {et~al.},\ }\bibfield  {title} {\enquote {\bibinfo {title}
  {Signatures of the $\text{Adler-Bell-Jackiw}$ chiral anomaly in a {Weyl}
  $\text{Fermion}$ semimetal},}\ }\href {\doibase 10.1038/ncomms10735}
  {\bibfield  {journal} {\bibinfo  {journal} {Nat. Commun.}\ }\textbf {\bibinfo
  {volume} {7}},\ \bibinfo {pages} {10735} (\bibinfo {year}
  {2016}{\natexlab{b}})}\BibitemShut {NoStop}%
\bibitem [{\citenamefont {He}\ \emph {et~al.}(2014)\citenamefont {He},
  \citenamefont {Hong}, \citenamefont {Dong}, \citenamefont {Pan},
  \citenamefont {Zhang}, \citenamefont {Zhang},\ and\ \citenamefont
  {Li}}]{He14prl}%
  \BibitemOpen
  \bibfield  {author} {\bibinfo {author} {\bibfnamefont {L.~P.}\ \bibnamefont
  {He}}, \bibinfo {author} {\bibfnamefont {X.~C.}\ \bibnamefont {Hong}},
  \bibinfo {author} {\bibfnamefont {J.~K.}\ \bibnamefont {Dong}}, \bibinfo
  {author} {\bibfnamefont {J.}~\bibnamefont {Pan}}, \bibinfo {author}
  {\bibfnamefont {Z.}~\bibnamefont {Zhang}}, \bibinfo {author} {\bibfnamefont
  {J.}~\bibnamefont {Zhang}}, \ and\ \bibinfo {author} {\bibfnamefont {S.~Y.}\
  \bibnamefont {Li}},\ }\bibfield  {title} {\enquote {\bibinfo {title} {Quantum
  transport evidence for the three-dimensional {Dirac} semimetal phase in
  $\text{Cd}_{3}\text{As}_{2}$},}\ }\href {\doibase
  10.1103/PhysRevLett.113.246402} {\bibfield  {journal} {\bibinfo  {journal}
  {Phys. Rev. Lett.}\ }\textbf {\bibinfo {volume} {113}},\ \bibinfo {pages}
  {246402} (\bibinfo {year} {2014})}\BibitemShut {NoStop}%
\bibitem [{\citenamefont {Liang}\ \emph {et~al.}(2015)\citenamefont {Liang},
  \citenamefont {Gibson}, \citenamefont {Ali}, \citenamefont {Liu},
  \citenamefont {Cava},\ and\ \citenamefont {Ong}}]{Liang15nmat}%
  \BibitemOpen
  \bibfield  {author} {\bibinfo {author} {\bibfnamefont {T.}~\bibnamefont
  {Liang}}, \bibinfo {author} {\bibfnamefont {Q.}~\bibnamefont {Gibson}},
  \bibinfo {author} {\bibfnamefont {M.~N.}\ \bibnamefont {Ali}}, \bibinfo
  {author} {\bibfnamefont {M.~H.}\ \bibnamefont {Liu}}, \bibinfo {author}
  {\bibfnamefont {R.~J.}\ \bibnamefont {Cava}}, \ and\ \bibinfo {author}
  {\bibfnamefont {N.~P.}\ \bibnamefont {Ong}},\ }\bibfield  {title} {\enquote
  {\bibinfo {title} {Ultrahigh mobility and giant magnetoresistance in the
  {Dirac} semimetal $\text{Cd}_3\text{As}_2$},}\ }\href
  {http://dx.doi.org/10.1038/nmat4143} {\bibfield  {journal} {\bibinfo
  {journal} {Nature Mater.}\ }\textbf {\bibinfo {volume} {14}},\ \bibinfo
  {pages} {280} (\bibinfo {year} {2015})}\BibitemShut {NoStop}%
\bibitem [{\citenamefont {Zhao}\ \emph {et~al.}(2015)\citenamefont {Zhao},
  \citenamefont {Liu}, \citenamefont {Zhang}, \citenamefont {Wang},
  \citenamefont {Wang}, \citenamefont {Lin}, \citenamefont {Xing},
  \citenamefont {Lu}, \citenamefont {Liu}, \citenamefont {Wang}, \citenamefont
  {Brombosz}, \citenamefont {Xiao}, \citenamefont {Jia}, \citenamefont {Xie},\
  and\ \citenamefont {Wang}}]{Zhao15prx}%
  \BibitemOpen
  \bibfield  {author} {\bibinfo {author} {\bibfnamefont {Y.~F.}\ \bibnamefont
  {Zhao}},  \emph {et~al.},\ }\bibfield  {title} {\enquote {\bibinfo {title}
  {Anisotropic $\text{Fermi}$ surface and quantum limit transport in high
  mobility three-dimensional {Dirac} semimetal $\text{Cd}_3\text{As}_2$},}\
  }\href {\doibase 10.1103/PhysRevX.5.031037} {\bibfield  {journal} {\bibinfo
  {journal} {Phys. Rev. X}\ }\textbf {\bibinfo {volume} {5}},\ \bibinfo {pages}
  {031037} (\bibinfo {year} {2015})}\BibitemShut {NoStop}%
\bibitem [{\citenamefont {Narayanan}\ \emph {et~al.}(2015)\citenamefont
  {Narayanan}, \citenamefont {Watson}, \citenamefont {Blake}, \citenamefont
  {Bruyant}, \citenamefont {Drigo}, \citenamefont {Chen}, \citenamefont
  {Prabhakaran}, \citenamefont {Yan}, \citenamefont {Felser}, \citenamefont
  {Kong}, \citenamefont {Canfield},\ and\ \citenamefont
  {Coldea}}]{Narayanan15prl}%
  \BibitemOpen
  \bibfield  {author} {\bibinfo {author} {\bibfnamefont {A.}~\bibnamefont
  {Narayanan}},  \emph {et~al.},\ }\bibfield  {title} {\enquote {\bibinfo
  {title} {Linear magnetoresistance caused by mobility fluctuations in
  $n$-doped $ \text{Cd}_3\text{As}_2$},}\ }\href {\doibase
  10.1103/PhysRevLett.114.117201} {\bibfield  {journal} {\bibinfo  {journal}
  {Phys. Rev. Lett.}\ }\textbf {\bibinfo {volume} {114}},\ \bibinfo {pages}
  {117201} (\bibinfo {year} {2015})}\BibitemShut {NoStop}%
\bibitem [{\citenamefont {Xiong}\ \emph {et~al.}(2015)\citenamefont {Xiong},
  \citenamefont {Kushwaha}, \citenamefont {Liang}, \citenamefont {Krizan},
  \citenamefont {Hirschberger}, \citenamefont {Wang}, \citenamefont {Cava},\
  and\ \citenamefont {Ong}}]{Xiong15sci}%
  \BibitemOpen
  \bibfield  {author} {\bibinfo {author} {\bibfnamefont {J.}~\bibnamefont
  {Xiong}}, \bibinfo {author} {\bibfnamefont {S.~K.}\ \bibnamefont {Kushwaha}},
  \bibinfo {author} {\bibfnamefont {T.}~\bibnamefont {Liang}}, \bibinfo
  {author} {\bibfnamefont {J.~W.}\ \bibnamefont {Krizan}}, \bibinfo {author}
  {\bibfnamefont {M.}~\bibnamefont {Hirschberger}}, \bibinfo {author}
  {\bibfnamefont {W.}~\bibnamefont {Wang}}, \bibinfo {author} {\bibfnamefont
  {R.~J.}\ \bibnamefont {Cava}}, \ and\ \bibinfo {author} {\bibfnamefont
  {N.~P.}\ \bibnamefont {Ong}},\ }\bibfield  {title} {\enquote {\bibinfo
  {title} {Evidence for the chiral anomaly in the {Dirac} semimetal
  $\text{Na}_3\text{Bi}$},}\ }\href
  {http://www.sciencemag.org/content/early/2015/09/02/science.aac6089.abstract}
  {\bibfield  {journal} {\bibinfo  {journal} {Science}\ }\textbf {\bibinfo
  {volume} {350}},\ \bibinfo {pages} {413} (\bibinfo {year}
  {2015})}\BibitemShut {NoStop}%
\bibitem [{\citenamefont {Li}\ \emph {et~al.}(2015)\citenamefont {Li},
  \citenamefont {Wang}, \citenamefont {Liu}, \citenamefont {Wang},
  \citenamefont {Liao},\ and\ \citenamefont {Yu}}]{LiCZ15nc}%
  \BibitemOpen
  \bibfield  {author} {\bibinfo {author} {\bibfnamefont {C.~Z.}\ \bibnamefont
  {Li}}, \bibinfo {author} {\bibfnamefont {L.~X.}\ \bibnamefont {Wang}},
  \bibinfo {author} {\bibfnamefont {H.~W.}\ \bibnamefont {Liu}}, \bibinfo
  {author} {\bibfnamefont {J.}~\bibnamefont {Wang}}, \bibinfo {author}
  {\bibfnamefont {Z.~M.}\ \bibnamefont {Liao}}, \ and\ \bibinfo {author}
  {\bibfnamefont {D.~P.}\ \bibnamefont {Yu}},\ }\bibfield  {title} {\enquote
  {\bibinfo {title} {Giant negative magnetoresistance induced by the chiral
  anomaly in individual $\text{Cd}_3\text{As}_2$ nanowires},}\ }\href {\doibase
  10.1038/ncomms10137} {\bibfield  {journal} {\bibinfo  {journal} {Nature
  Commun.}\ }\textbf {\bibinfo {volume} {6}},\ \bibinfo {pages} {10137}
  (\bibinfo {year} {2015})}\BibitemShut {NoStop}%
\bibitem [{\citenamefont {Li}\ \emph {et~al.}(2016)\citenamefont {Li},
  \citenamefont {He}, \citenamefont {Lu}, \citenamefont {Zhang}, \citenamefont
  {Liu}, \citenamefont {Ma}, \citenamefont {Fan}, \citenamefont {Shen},\ and\
  \citenamefont {Wang}}]{LiH16nc}%
  \BibitemOpen
  \bibfield  {author} {\bibinfo {author} {\bibfnamefont {H.}~\bibnamefont
  {Li}}, \bibinfo {author} {\bibfnamefont {H.~T.}\ \bibnamefont {He}}, \bibinfo
  {author} {\bibfnamefont {H.~Z.}\ \bibnamefont {Lu}}, \bibinfo {author}
  {\bibfnamefont {H.~C.}\ \bibnamefont {Zhang}}, \bibinfo {author}
  {\bibfnamefont {H.~C.}\ \bibnamefont {Liu}}, \bibinfo {author} {\bibfnamefont
  {R.}~\bibnamefont {Ma}}, \bibinfo {author} {\bibfnamefont {Z.~Y.}\
  \bibnamefont {Fan}}, \bibinfo {author} {\bibfnamefont {S.~Q.}\ \bibnamefont
  {Shen}}, \ and\ \bibinfo {author} {\bibfnamefont {J.~N.}\ \bibnamefont
  {Wang}},\ }\bibfield  {title} {\enquote {\bibinfo {title} {Negative
  magnetoresistance in {Dirac} semimetal $\text{Cd}_3\text{As}_2$},}\ }\href
  {\doibase 10.1038/ncomms10301} {\bibfield  {journal} {\bibinfo  {journal}
  {Nature Commun.}\ }\textbf {\bibinfo {volume} {7}},\ \bibinfo {pages} {10301}
  (\bibinfo {year} {2016})}\BibitemShut {NoStop}%
\bibitem [{\citenamefont {Zhang}\ \emph {et~al.}(2017)\citenamefont {Zhang},
  \citenamefont {Zhang}, \citenamefont {Wang}, \citenamefont {Liu},
  \citenamefont {Chen}, \citenamefont {Lu}, \citenamefont {Liang},
  \citenamefont {Cao}, \citenamefont {Yuan}, \citenamefont {Tang},
  \citenamefont {Li}, \citenamefont {Zhou}, \citenamefont {Gu}, \citenamefont
  {Wu}, \citenamefont {Zou},\ and\ \citenamefont {Xiu}}]{ZhangC17nc}%
  \BibitemOpen
  \bibfield  {author} {\bibinfo {author} {\bibfnamefont {C.}~\bibnamefont
  {Zhang}},  \emph {et~al.},\ }\bibfield  {title} {\enquote {\bibinfo {title}
  {Room-temperature chiral charge pumping in {Dirac} semimetals},}\ }\href
  {\doibase 10.1038/ncomms13741} {\bibfield  {journal} {\bibinfo  {journal}
  {Nat. Commun.}\ }\textbf {\bibinfo {volume} {8}},\ \bibinfo {pages} {13741}
  (\bibinfo {year} {2017})}\BibitemShut {NoStop}%
\bibitem [{\citenamefont {Uchida}\ \emph {et~al.}(2017)\citenamefont {Uchida},
  \citenamefont {Nakazawa}, \citenamefont {Nishihaya}, \citenamefont {Akiba},
  \citenamefont {Kriener}, \citenamefont {Kozuka}, \citenamefont {Miyake},
  \citenamefont {Taguchi}, \citenamefont {Tokunaga}, \citenamefont {Nagaosa},
  \citenamefont {Tokura},\ and\ \citenamefont
  {Kawasaki}}]{Uchida17MarchMeeting}%
  \BibitemOpen
  \bibfield  {author} {\bibinfo {author} {\bibfnamefont {M.}~\bibnamefont
  {Uchida}},  \emph {et~al.},\ }\bibfield  {title} {\enquote {\bibinfo {title}
  {Quantum {Hall} effect in {Cd}$_3${As}$_2$ films},}\ }\href
  {https://meetings.aps.org/Meeting/MAR17/Session/A44.5} {\bibfield  {journal}
  {\bibinfo  {journal} {APS March Meeting A44.00005}\ } (\bibinfo {year}
  {2017})}\BibitemShut {NoStop}%
\bibitem [{\citenamefont {Shen}(2012)}]{Shen12book}%
  \BibitemOpen
  \bibfield  {author} {\bibinfo {author} {\bibfnamefont {S.-Q.}\ \bibnamefont
  {Shen}},\ }\href@noop {} {\emph {\bibinfo {title} {Topological Insulators}}}\
  (\bibinfo  {publisher} {Springer-Verlag},\ \bibinfo {address} {Berlin
  Heidelberg},\ \bibinfo {year} {2012})\BibitemShut {NoStop}%
\bibitem [{\citenamefont {Okugawa}\ and\ \citenamefont
  {Murakami}(2014)}]{Okugawa14prb}%
  \BibitemOpen
  \bibfield  {author} {\bibinfo {author} {\bibfnamefont {R.}~\bibnamefont
  {Okugawa}}\ and\ \bibinfo {author} {\bibfnamefont {S.}~\bibnamefont
  {Murakami}},\ }\bibfield  {title} {\enquote {\bibinfo {title} {Dispersion of
  fermi arcs in weyl semimetals and their evolutions to dirac cones},}\ }\href
  {\doibase 10.1103/PhysRevB.89.235315} {\bibfield  {journal} {\bibinfo
  {journal} {Phys. Rev. B}\ }\textbf {\bibinfo {volume} {89}},\ \bibinfo
  {pages} {235315} (\bibinfo {year} {2014})}\BibitemShut {NoStop}%
\bibitem [{\citenamefont {Lu}\ \emph {et~al.}(2015)\citenamefont {Lu},
  \citenamefont {Zhang},\ and\ \citenamefont {Shen}}]{Lu15Weyl-shortrange}%
  \BibitemOpen
  \bibfield  {author} {\bibinfo {author} {\bibfnamefont {H.~Z.}\ \bibnamefont
  {Lu}}, \bibinfo {author} {\bibfnamefont {S.~B.}\ \bibnamefont {Zhang}}, \
  and\ \bibinfo {author} {\bibfnamefont {S.~Q.}\ \bibnamefont {Shen}},\
  }\bibfield  {title} {\enquote {\bibinfo {title} {High-field
  magnetoconductivity of topological semimetals with short-range potential},}\
  }\href {\doibase 10.1103/PhysRevB.92.045203} {\bibfield  {journal} {\bibinfo
  {journal} {Phys. Rev. B}\ }\textbf {\bibinfo {volume} {92}},\ \bibinfo
  {pages} {045203} (\bibinfo {year} {2015})}\BibitemShut {NoStop}%
\bibitem [{\citenamefont {Lu}\ and\ \citenamefont {Shen}(2017)}]{Lu17fop}%
  \BibitemOpen
  \bibfield  {author} {\bibinfo {author} {\bibfnamefont {H.-Z.}\ \bibnamefont
  {Lu}}\ and\ \bibinfo {author} {\bibfnamefont {S.-Q.}\ \bibnamefont {Shen}},\
  }\bibfield  {title} {\enquote {\bibinfo {title} {Quantum transport in
  topological semimetals under magnetic fields},}\ }\href {\doibase
  10.1007/s11467-016-0609-y} {\bibfield  {journal} {\bibinfo  {journal} {Front.
  Phys.}\ }\textbf {\bibinfo {volume} {12}},\ \bibinfo {pages} {127201}
  (\bibinfo {year} {2017})}\BibitemShut {NoStop}%
\bibitem [{\citenamefont {Lu}\ \emph {et~al.}(2010)\citenamefont {Lu},
  \citenamefont {Shan}, \citenamefont {Yao}, \citenamefont {Niu},\ and\
  \citenamefont {Shen}}]{Lu10prb}%
  \BibitemOpen
  \bibfield  {author} {\bibinfo {author} {\bibfnamefont {H.~Z.}\ \bibnamefont
  {Lu}}, \bibinfo {author} {\bibfnamefont {W.~Y.}\ \bibnamefont {Shan}},
  \bibinfo {author} {\bibfnamefont {W.}~\bibnamefont {Yao}}, \bibinfo {author}
  {\bibfnamefont {Q.}~\bibnamefont {Niu}}, \ and\ \bibinfo {author}
  {\bibfnamefont {S.~Q.}\ \bibnamefont {Shen}},\ }\bibfield  {title} {\enquote
  {\bibinfo {title} {Massive {Dirac} fermions and spin physics in an ultrathin
  film of topological insulator},}\ }\href {\doibase
  10.1103/PhysRevB.81.115407} {\bibfield  {journal} {\bibinfo  {journal} {Phys.
  Rev. B}\ }\textbf {\bibinfo {volume} {81}},\ \bibinfo {pages} {115407}
  (\bibinfo {year} {2010})}\BibitemShut {NoStop}%
\bibitem [{\citenamefont {Shan}\ \emph {et~al.}(2010)\citenamefont {Shan},
  \citenamefont {Lu},\ and\ \citenamefont {Shen}}]{Shan11njp}%
  \BibitemOpen
  \bibfield  {author} {\bibinfo {author} {\bibfnamefont {W.-Y.}\ \bibnamefont
  {Shan}}, \bibinfo {author} {\bibfnamefont {H.-Z.}\ \bibnamefont {Lu}}, \ and\
  \bibinfo {author} {\bibfnamefont {S.-Q.}\ \bibnamefont {Shen}},\ }\bibfield
  {title} {\enquote {\bibinfo {title} {Effective continuous model for surface
  states and thin films of three-dimensional topological insulators},}\ }\href
  {http://stacks.iop.org/1367-2630/12/i=4/a=043048} {\bibfield  {journal}
  {\bibinfo  {journal} {New J. Phys.}\ }\textbf {\bibinfo {volume} {12}},\
  \bibinfo {pages} {043048} (\bibinfo {year} {2010})}\BibitemShut {NoStop}%
\bibitem [{Sup()}]{Supp}%
  \BibitemOpen
  \href@noop {} {\bibinfo  {journal} {Supplemental Material}\ }\BibitemShut
  {NoStop}%
\bibitem [{\citenamefont {Rosenberg}\ \emph {et~al.}(2010)\citenamefont
  {Rosenberg}, \citenamefont {Guo},\ and\ \citenamefont
  {Franz}}]{Rosenberg10prb}%
  \BibitemOpen
\bibfield  {journal} {  }\bibfield  {author} {\bibinfo {author} {\bibfnamefont
  {G.}~\bibnamefont {Rosenberg}}, \bibinfo {author} {\bibfnamefont {H.-M.}\
  \bibnamefont {Guo}}, \ and\ \bibinfo {author} {\bibfnamefont
  {M.}~\bibnamefont {Franz}},\ }\bibfield  {title} {\enquote {\bibinfo {title}
  {Wormhole effect in a strong topological insulator},}\ }\href {\doibase
  10.1103/PhysRevB.82.041104} {\bibfield  {journal} {\bibinfo  {journal} {Phys.
  Rev. B}\ }\textbf {\bibinfo {volume} {82}},\ \bibinfo {pages} {041104}
  (\bibinfo {year} {2010})}\BibitemShut {NoStop}%
\bibitem [{\citenamefont {Gusynin}\ and\ \citenamefont
  {Sharapov}(2005)}]{Gusynin05prl}%
  \BibitemOpen
  \bibfield  {author} {\bibinfo {author} {\bibfnamefont {V.~P.}\ \bibnamefont
  {Gusynin}}\ and\ \bibinfo {author} {\bibfnamefont {S.~G.}\ \bibnamefont
  {Sharapov}},\ }\bibfield  {title} {\enquote {\bibinfo {title} {Unconventional
  integer quantum {Hall} effect in graphene},}\ }\href {\doibase
  10.1103/PhysRevLett.95.146801} {\bibfield  {journal} {\bibinfo  {journal}
  {Phys. Rev. Lett.}\ }\textbf {\bibinfo {volume} {95}},\ \bibinfo {pages}
  {146801} (\bibinfo {year} {2005})}\BibitemShut {NoStop}%
\bibitem [{\citenamefont {Zyuzin}\ and\ \citenamefont
  {Burkov}(2011)}]{Zyuzin11prb}%
  \BibitemOpen
  \bibfield  {author} {\bibinfo {author} {\bibfnamefont {A.~A.}\ \bibnamefont
  {Zyuzin}}\ and\ \bibinfo {author} {\bibfnamefont {A.~A.}\ \bibnamefont
  {Burkov}},\ }\bibfield  {title} {\enquote {\bibinfo {title} {Thin topological
  insulator film in a perpendicular magnetic field},}\ }\href {\doibase
  10.1103/PhysRevB.83.195413} {\bibfield  {journal} {\bibinfo  {journal} {Phys.
  Rev. B}\ }\textbf {\bibinfo {volume} {83}},\ \bibinfo {pages} {195413}
  (\bibinfo {year} {2011})}\BibitemShut {NoStop}%
\bibitem [{\citenamefont {Zhang}\ \emph {et~al.}(2014)\citenamefont {Zhang},
  \citenamefont {Zhang},\ and\ \citenamefont {Shen}}]{ZhangSB14prb}%
  \BibitemOpen
  \bibfield  {author} {\bibinfo {author} {\bibfnamefont {S.~B.}\ \bibnamefont
  {Zhang}}, \bibinfo {author} {\bibfnamefont {Y.~Y.}\ \bibnamefont {Zhang}}, \
  and\ \bibinfo {author} {\bibfnamefont {S.~Q.}\ \bibnamefont {Shen}},\
  }\bibfield  {title} {\enquote {\bibinfo {title} {Robustness of quantum spin
  {Hall} effect in an external magnetic field},}\ }\href {\doibase
  10.1103/PhysRevB.90.115305} {\bibfield  {journal} {\bibinfo  {journal} {Phys.
  Rev. B}\ }\textbf {\bibinfo {volume} {90}},\ \bibinfo {pages} {115305}
  (\bibinfo {year} {2014})}\BibitemShut {NoStop}%
\bibitem [{\citenamefont {Zhang}\ \emph {et~al.}(2015)\citenamefont {Zhang},
  \citenamefont {Lu},\ and\ \citenamefont {Shen}}]{ZhangSB15srep}%
  \BibitemOpen
  \bibfield  {author} {\bibinfo {author} {\bibfnamefont {S.~B.}\ \bibnamefont
  {Zhang}}, \bibinfo {author} {\bibfnamefont {H.~Z.}\ \bibnamefont {Lu}}, \
  and\ \bibinfo {author} {\bibfnamefont {S.~Q.}\ \bibnamefont {Shen}},\
  }\bibfield  {title} {\enquote {\bibinfo {title} {Edge states and integer
  quantum {Hall} effect in topological insulator thin films},}\ }\href
  {\doibase 10.1038/srep13277} {\bibfield  {journal} {\bibinfo  {journal} {Sci.
  Rep.}\ }\textbf {\bibinfo {volume} {5}},\ \bibinfo {pages} {13277} (\bibinfo
  {year} {2015})}\BibitemShut {NoStop}%
\bibitem [{\citenamefont {Pertsova}\ \emph {et~al.}(2016)\citenamefont
  {Pertsova}, \citenamefont {Canali},\ and\ \citenamefont
  {MacDonald}}]{Pertsova16prb}%
  \BibitemOpen
  \bibfield  {author} {\bibinfo {author} {\bibfnamefont {A.}~\bibnamefont
  {Pertsova}}, \bibinfo {author} {\bibfnamefont {C.~M.}\ \bibnamefont
  {Canali}}, \ and\ \bibinfo {author} {\bibfnamefont {A.~H.}\ \bibnamefont
  {MacDonald}},\ }\bibfield  {title} {\enquote {\bibinfo {title} {Quantum
  {Hall} edge states in topological insulator nanoribbons},}\ }\href {\doibase
  10.1103/PhysRevB.94.121409} {\bibfield  {journal} {\bibinfo  {journal} {Phys.
  Rev. B}\ }\textbf {\bibinfo {volume} {94}},\ \bibinfo {pages} {121409}
  (\bibinfo {year} {2016})}\BibitemShut {NoStop}%
\bibitem [{\citenamefont {Jain}(2007)}]{Jain07book}%
  \BibitemOpen
  \bibfield  {author} {\bibinfo {author} {\bibfnamefont {J.~K.}\ \bibnamefont
  {Jain}},\ }\href@noop {} {\emph {\bibinfo {title} {Composite fermions}}}\
  (\bibinfo  {publisher} {Cambridge University Press},\ \bibinfo {year}
  {2007})\BibitemShut {NoStop}%
\bibitem [{\citenamefont {Laughlin}(1981)}]{Laughlin81prb}%
  \BibitemOpen
  \bibfield  {author} {\bibinfo {author} {\bibfnamefont {R.~B.}\ \bibnamefont
  {Laughlin}},\ }\bibfield  {title} {\enquote {\bibinfo {title} {Quantized
  {Hall} conductivity in two dimensions},}\ }\href {\doibase
  10.1103/PhysRevB.23.5632} {\bibfield  {journal} {\bibinfo  {journal} {Phys.
  Rev. B}\ }\textbf {\bibinfo {volume} {23}},\ \bibinfo {pages} {5632}
  (\bibinfo {year} {1981})}\BibitemShut {NoStop}%
\bibitem [{\citenamefont {Cano}\ \emph {et~al.}(2017)\citenamefont {Cano},
  \citenamefont {Bradlyn}, \citenamefont {Wang}, \citenamefont {Hirschberger},
  \citenamefont {Ong},\ and\ \citenamefont {Bernevig}}]{Cano17prbrc}%
  \BibitemOpen
  \bibfield  {author} {\bibinfo {author} {\bibfnamefont {J.}~\bibnamefont
  {Cano}}, \bibinfo {author} {\bibfnamefont {B.}~\bibnamefont {Bradlyn}},
  \bibinfo {author} {\bibfnamefont {Z.}~\bibnamefont {Wang}}, \bibinfo {author}
  {\bibfnamefont {M.}~\bibnamefont {Hirschberger}}, \bibinfo {author}
  {\bibfnamefont {N.~P.}\ \bibnamefont {Ong}}, \ and\ \bibinfo {author}
  {\bibfnamefont {B.~A.}\ \bibnamefont {Bernevig}},\ }\bibfield  {title}
  {\enquote {\bibinfo {title} {Chiral anomaly factory: Creating {Weyl} fermions
  with a magnetic field},}\ }\href {\doibase 10.1103/PhysRevB.95.161306}
  {\bibfield  {journal} {\bibinfo  {journal} {Phys. Rev. B}\ }\textbf {\bibinfo
  {volume} {95}},\ \bibinfo {pages} {161306} (\bibinfo {year}
  {2017})}\BibitemShut {NoStop}%
\bibitem [{\citenamefont {Jeon}\ \emph {et~al.}(2014)\citenamefont {Jeon},
  \citenamefont {Zhou}, \citenamefont {Gyenis}, \citenamefont {Feldman},
  \citenamefont {Kimchi}, \citenamefont {Potter}, \citenamefont {Gibson},
  \citenamefont {Cava}, \citenamefont {Vishwanath},\ and\ \citenamefont
  {Yazdani}}]{Jeon14nmat}%
  \BibitemOpen
  \bibfield  {author} {\bibinfo {author} {\bibfnamefont {S.}~\bibnamefont
  {Jeon}}, \bibinfo {author} {\bibfnamefont {B.~B.}\ \bibnamefont {Zhou}},
  \bibinfo {author} {\bibfnamefont {A.}~\bibnamefont {Gyenis}}, \bibinfo
  {author} {\bibfnamefont {B.~E.}\ \bibnamefont {Feldman}}, \bibinfo {author}
  {\bibfnamefont {I.}~\bibnamefont {Kimchi}}, \bibinfo {author} {\bibfnamefont
  {A.~C.}\ \bibnamefont {Potter}}, \bibinfo {author} {\bibfnamefont {Q.~D.}\
  \bibnamefont {Gibson}}, \bibinfo {author} {\bibfnamefont {R.~J.}\
  \bibnamefont {Cava}}, \bibinfo {author} {\bibfnamefont {A.}~\bibnamefont
  {Vishwanath}}, \ and\ \bibinfo {author} {\bibfnamefont {A.}~\bibnamefont
  {Yazdani}},\ }\bibfield  {title} {\enquote {\bibinfo {title} {Landau
  quantization and quasiparticle interference in the three-dimensional {Dirac}
  semimetal $\text{Cd}_3\text{As}_2$},}\ }\href
  {http://dx.doi.org/10.1038/nmat4023} {\bibfield  {journal} {\bibinfo
  {journal} {Nature Mater.}\ }\textbf {\bibinfo {volume} {13}},\ \bibinfo
  {pages} {851} (\bibinfo {year} {2014})}\BibitemShut {NoStop}%
\bibitem [{\citenamefont {Kargarian}\ \emph {et~al.}(2016)\citenamefont
  {Kargarian}, \citenamefont {Randeria},\ and\ \citenamefont
  {Lu}}]{Kargarian16pnas}%
  \BibitemOpen
  \bibfield  {author} {\bibinfo {author} {\bibfnamefont {M.}~\bibnamefont
  {Kargarian}}, \bibinfo {author} {\bibfnamefont {M.}~\bibnamefont {Randeria}},
  \ and\ \bibinfo {author} {\bibfnamefont {Y.-M.}\ \bibnamefont {Lu}},\
  }\bibfield  {title} {\enquote {\bibinfo {title} {Are the surface {Fermi} arcs
  in {Dirac} semimetals topologically protected?}}\ }\href {\doibase
  10.1073/pnas.1524787113} {\bibfield  {journal} {\bibinfo  {journal}
  {Proceedings of the National Academy of Sciences}\ }\textbf {\bibinfo
  {volume} {113}},\ \bibinfo {pages} {8648} (\bibinfo {year}
  {2016})}\BibitemShut {NoStop}%
\end{thebibliography}


\end{document}